\documentclass[aps,superscriptaddress,showpacs,showkeys,nofootinbib,floatfix]{revtex4}
\usepackage{graphicx,epsfig,wrapfig,bbm}
\usepackage{epstopdf}
\usepackage{bm}
\usepackage{amssymb,amsmath,slashed,slashbox,multirow,amsfonts,latexsym}
\usepackage[normalem]{ulem} 
\renewcommand\sout{\bgroup \color[rgb]{0.55,0.00,0.99} \ULdepth=-.5ex \ULset}

\usepackage{color}
\usepackage{dsfont}
\usepackage{pdfpages}
\definecolor{darkgreen}{rgb}{0,0.65,0}

\definecolor{green}{rgb}{0,.5,0}

\definecolor{orange}{rgb}{1,0.5,0}

\newcommand{\be}{\begin{equation}}
\newcommand{\ee}{\end{equation}}
\newcommand{\ba}{\begin{eqnarray}}
\newcommand{\ea}{\end{eqnarray}}
\newcommand{\la}{\langle}
\newcommand{\ra}{\rangle}
\newcommand{\di}{\mathrm{d}}
\newcommand{\ud}{\mathrm{d}}

\newcommand{\uTr}{\mathrm{Tr}}

\newcommand{\uslash}{/\!\!\!}

\newcommand{\up}{\boldsymbol{p}}

\newcommand{\uzero}{\boldsymbol{0}}

\newlength\savedwidth

\begin{document}
%
\newcommand*{\SLAC}{SLAC National Accelerator Laboratory, Stanford 
  University, Menlo Park, California 94025 USA}\affiliation{\SLAC}
\newcommand*{\ULg}{IFPA, AGO Department, Universit\'e de Li\`ege, 
  Sart-Tilman, 4000 Li\`ege, Belgium}\affiliation{\ULg}
\newcommand*{\Pavia}{Dipartimento di Fisica, 
  Universit\`a degli Studi di Pavia, Pavia, Italy}\affiliation{\Pavia}
\newcommand*{\INFN}{Istituto Nazionale di Fisica Nucleare, 
  Sezione di Pavia, Pavia, Italy}\affiliation{\INFN}
\newcommand*{\UConn}{Department of Physics, University of Connecticut, 
  Storrs, CT 06269, USA}\affiliation{\UConn}
\title{
Transverse-momentum dependent parton distribution functions \\
  beyond leading twist in quark models}
\author{C.~Lorc\'e}\affiliation{\SLAC}\affiliation{\ULg}
\author{B.~Pasquini}\affiliation{\Pavia}\affiliation{\INFN}
\author{P.~Schweitzer}\affiliation{\UConn}
\vspace{0.5in}
\begin{abstract}
    Higher-twist transverse momentum dependent parton distribution
    functions (TMDs) are a valuable probe of the quark-gluon dynamics in 
    the nucleon, and play a vital role for the explanation of 
    sizable azimuthal asymmetries in hadron production from unpolarized 
    and polarized deep-inelastic lepton-nucleon scattering observed 
    in experiments at CERN, DESY and Jefferson Lab. The associated 
    observables are challenging to interpret, and still await a complete 
    theoretical explanation, which makes guidance from models valuable.
    In this work we establish the formalism to describe unpolarized 
    higher-twist TMDs in the light-front framework based on a Fock-space 
    expansion of the nucleon state in terms of free on-shell parton states. 
    We derive general expressions and present numerical results in 
    a practical realization of this picture provided by the light-front
    constituent quark model. We review several other popular quark 
    model approaches including free quark ensemble, bag, spectator
    and chiral quark-soliton model. 
    We discuss how higher-twist TMDs are described in these models, 
    and obtain results for several TMDs not 
    discussed previously in literature. 
    This study 
    contributes to the understanding of non-perturbative properties 
    of subleading twist TMDs.
    The results from the light-front constituent 
    quark model are also compared to available phenomenological
    information, showing a satisfactory agreement. 
\end{abstract}
\pacs{
      12.39.Ki, 
      13.60.Hb, 
      13.85.Qk} 
\keywords{quark-gluon structure, higher twist,
          transverse momentum dependent distribution functions}
\maketitle
\section{Introduction}
\label{Sec-1:introduction}

Azimuthal (spin) asymmetries in semi-inclusive deep-inelastic scattering 
(SIDIS) due to transverse parton momenta~\cite{Kotzinian:1994dv,Mulders:1995dh}
can be classified as unsuppressed leading-twist (twist-2) and power-suppressed 
subleading-twist  (twist-3) effects in the sense of the 
``working twist definition'' of  Ref.~\cite{Jaffe:1996zw}. 
The theoretical description of leading-twist effects is cleaner, 
and clear experimental evidence is available, see~\cite{Burkardt:2008jw,Barone:2010zz,Aidala:2012mv} for reviews. 
However, the first measurements of such asymmetries in SIDIS in 
unpolarized case by EMC~\cite{Aubert:1983cz,Arneodo:1986cf} 
or with longitudinally polarized targets by HERMES~\cite{Avakian:1999rr,Airapetian:1999tv,Airapetian:2001eg,Airapetian:2002mf} 
unexpectedly revealed larger effects at twist-3 level than at twist-2
in the fixed-target kinematics of these experiments. 
Further data on twist-3 SIDIS effects (including preliminary results) was
reported in Refs.~\cite{Avakian:2003pk,Airapetian:2005jc,Airapetian:2006rx,
Mkrtchyan:2007sr,Osipenko:2008aa,Avakian:2010ae,Aghasyan:2011ha,Gohn:2014zbz,
Kotzinian:2007uv,Parsamyan:2013ug}.

SIDIS is a rich source of information on the nucleon structure
including subleading-twist effects. However, in a tree-level factorization 
approach, twist-3 SIDIS observables receive 4 (or 6) contributions due to
twist-3 (or twist-2) transverse momentum dependent parton distribution 
functions (TMDs) convoluted with twist-2 (or twist-3) transverse momentum 
dependent fragmentation functions~\cite{Bacchetta:2006tn}.
This makes the theoretical interpretation of data challenging,
and motivates model studies to help to clarify the underlying physics.
The important impact of model studies for the understanding of TMDs was 
reviewed in~\cite{Metz:2004ya}. 
Model calculations also indicate that the status of TMD factorization in 
SIDIS beyond leading twist is not yet fully clarified~\cite{Gamberg:2006ru}. Information on collinear twist-3 parton distribution functions 
is limited to $g_T^q(x)$ accessed in polarized DIS, see~\cite{Accardi:2009au}
for an overview. The interference fragmentation function approach
based on collinear factorization offers a way to access further twist-3 
parton distribution functions in a collinear factorization~\cite{Bacchetta:2003vn}. 
A first extraction of one of these functions, namely $e^q(x)$,
using this framework was recently reported in Ref.~\cite{Courtoy:2014ixa}.

Higher-twist TMDs can in general be decomposed in contributions
from leading-twist, current quark mass terms and pure interaction-dependent (``tilde'') terms. This is accomplished by employing
equations of motion (EOM) and reveals that tilde-terms are not parton 
densities but quark-gluon correlation functions. 
Neglecting the tilde- and mass terms is sometimes referred 
to as Wandzura-Wilczek (WW) approximation~\cite{Wandzura:1977qf}. 
This step can be helpful in phenomenology to disentangle the 
many contributions to twist-3 SIDIS 
observables~\cite{DeSanctis:2000fh,Ma:2000ip,Efremov:2001cz,Efremov:2002ut}, 
and can in certain cases be a numerically useful 
approximation~\cite{Avakian:2007mv,Accardi:2009au}.
But it removes the richness of the largely unexplored 
but attractive non-perturbative physics of quark-gluon 
correlations. Precisely this is an important motivation to study 
subleading-twist effects~\cite{Jaffe:1983hp,Jaffe:1989xx}.

Higher-twist TMDs and parton distribution functions of quarks are 
expressed in terms of hadronic matrix elements of bilinear quark-field 
correlators of the type $\la h|\overline\psi(0)\Gamma\psi(z)|h\ra$, which 
makes them amenable to studies in quark models~\cite{Jaffe:1991ra}, defined in the following as models without explicit gauge-field degrees of freedom.
Quark models with interactions allow one, in principle, to model 
also the interaction-dependent tilde-terms. 
Quark models have been shown to give a useful description of 
leading-twist TMDs and related SIDIS observables, provided one
applies them carefully within their range of applicability.
Much less is known about higher-twist TMDs, and important 
questions emerge.
What precisely can we learn from quark models? To what extent can 
quark models give estimates for higher-twist TMDs? And how useful 
are such estimates phenomenologically? 
This work will not provide an extensive answer to these complex 
questions. But it will, as we hope, shed new light on the applicability of 
quark models to TMDs beyond leading twist. In this work we will limit 
ourselves to unpolarized higher-twist TMDs. Earlier work in this sector was 
presented in~\cite{Jaffe:1991ra,Signal:1996ct,
Jakob:1997wg,Schweitzer:2003uy,Wakamatsu:2003uu,Avakian:2010br}.

The specific goals of this work are as follows. After a brief introduction 
on unpolarized TMDs in Sec.~\ref{Sec-2:counting-structures}, we will work 
out in Sec.~\ref{Sec-3:eom-general-strategy} a general approach to derive 
unique decompositions 
of subleading-twist TMDs into 
twist-2 parts and mass terms by making use of the free EOM, where
tilde-terms are absent. In the subsequent sections we will generalize 
this formalism to include interactions in specific quark models,
which will give rise to tilde-terms.

In Secs.~\ref{section-51}--\ref{Sec-7:eom-in-interacting-models} 
we will discuss several quark models, starting with the ensemble of free 
quarks~\cite{Tangerman:1994eh}, a prototype for parton-model frameworks 
where interactions are absent. When discussing interacting models, we 
will include the spectator~\cite{Jakob:1997wg}, 
chiral quark-soliton~\cite{Schweitzer:2003uy,Wakamatsu:2003uu} 
and bag~\cite{Avakian:2010br} models,
and investigate how interaction-dependent tilde-terms arise in those models. 
Hereby we will not only review available results, but also present new 
results not discussed previously in the literature. We will also derive a 
so-called Lorentz-invariance relation (LIR) among unpolarized TMDs valid 
in frameworks without gauge degrees of freedom, \emph{i.e.} also in 
quark models.
We will use the LIR to test the theoretical consistency of the 
model frameworks.

A central part of this work is Sect.~\ref{section-52}. 
Here we will extend the light-front constituent quark model approach
(LFCQM), which was used in the past to study leading-twist TMDs, to 
the description of higher-twist TMDs.
This model in some sense exhibits features of both free and interacting 
quark models. In fact, we will find that some (not all) of the relations 
among TMDs derived from free EOM hold, which can be traced back to the fact 
that this approach is based on a light-front Fock-state expansion of the 
nucleon state in terms of on-shell parton states obeying the free EOM.
However, we also find that the LIR is not supported in the LFCQM. 
Technically this is because the single quarks are on-shell, but the 
three-quark state they form is not, with the off-shellness introduced by the 
non-perturbative bound state information encoded in the nucleon wave-function.
The deeper and more general reasons for the non-compliance with the LIR 
can be traced back to generic issues with the conservation of the 
minus-component of the electromagnetic current in light-front 
approaches, which requires the inclusion of higher Fock states 
not accounted for in this approach.

The paper is rounded up by Sec.~\ref{Sec-X:numerical-results}, where we 
will present and compare the numerical results from the quark models. 
We will also confront predictions from the LFCQM to available results 
from phenomenology on $e^q(x)$. After the conclusions in 
Sec.~\ref{Sec-8:conclusions}, we will present Appendices 
with technical details.

\section{TMD\lowercase{s} and equations of motion relations}
\label{Sec-2:counting-structures}

Quark and antiquark TMDs for flavor $q$ are defined in QCD in terms of 
quark correlators of the type
\be\label{Eq:correlator}
   \Phi^q_{ij}(P,p,S;\mbox{path}) = \int \frac{\di^4z}{(2\pi)^4} \, 
   e^{i p \cdot z} \, \la P,S  |  \overline{\psi}_j(0) 
   \, {\cal W}(0,z;\mbox{path}) \, \psi_i(z)  |  P,S\ra ,
\ee
where $P$ ($S$) denotes four-momentum (polarization vector) of the nucleon,
and $p$ is the four-momentum of the quark. TMDs are given by such 
correlators integrated over $p^-$ with $p^+=xP^+$. 
Factorization theorems dictate (for $p^-$-integrated correlators) 
the process-dependent ``path'' indicated in Eq.~\eqref{Eq:correlator} 
along which appropriate Wilson lines connect the bi-local quark field 
operators~\cite{Collins-book}. (Our notation is
$p^\pm = \tfrac{1}{\sqrt{2}}(p^0 \pm p^3)$ and $\bm{p}_T=(p^1,p^2)$
with $p_T\equiv|\bm{p}_T|$. The metric is such that for instance
$a\cdot b$ = $a^+b^-+a^-b^+-\bm{a}_T\cdot\bm{b}_T$ 
and $\di^4 z = \di z^+\di z^-\di^2 z_T$. For brevity we do not 
indicate the scale dependence of the correlators and TMDs, and often 
omit the flavor index $q$ on the quark fields $\psi\equiv \psi^q$.)

In order to count independent structures, one decomposes the
correlator in terms of scalar ``amplitudes'' multiplied by independent 
Lorentz structures allowed by the symmetries of the strong interactions
and constructed from the four-vectors $P$, $S$, 
$p$~\cite{Tangerman:1994bb,Tangerman:1994eh} and a (near-)lightlike 
four-vector $n$~\cite{Goeke:2005hb} 
which characterizes the path of the Wilson line 
(actually, the situation is more complex than that~\cite{Accardi:2009au}, 
but this does not change the general conclusion~\cite{Goeke:2005hb}).
In QCD one has 32 independent amplitudes: $A^q_i$ with $1\le i \le 12$ 
and $B^q_j$ with $1\le j \le 20$~\cite{Goeke:2005hb}. There are also 32 TMDs: 
namely 8 at leading twist, 16 at twist-3, and 8 (more academic) at twist-4. 
Thus, one ends up with as many TMDs as amplitudes, and there are 
\emph{a priori} no relations among  TMDs, unless one resorts to
approximations such as the above-mentioned WW approximations.

What distinguishes the $A^q_i$ and $B^q_i$ is that the former multiply
Lorentz structures made from  $P$, $S$, $p$ only, while the latter
explicitly include also the vector $n$ characterizing the gauge-link.
Therefore, in quark models (with no gauge fields) all the $B^q_i$ amplitudes 
are absent. Moreover, the amplitudes $A^q_4$, $A^q_5$, $A^q_{12}$ are ``naively 
$\mathsf{T}$-odd'' which is allowed 
in QCD~\cite{Sivers:1989cc,Boer:1997nt,Brodsky:2002cx,Collins:2002kn},
but forbidden in quark models~\cite{Pobylitsa:2002fr}.
Thus, in quark models up to twist-3, one has 9 amplitudes describing 14 
$\mathsf{T}$-even TMDs, out of which 6 (8) are twist-2 (twist-3). This 
implies the existence of 5 ``Lorentz-invariance relations'' (LIRs)
among T-even TMDs~\cite{Tangerman:1994bb,Tangerman:1994eh} which
must hold in quark models~\cite{Avakian:2007mv}, but are not valid in QCD~\cite{Goeke:2003az} due to the presence of 
$B^q_i$-amplitudes~\cite{Goeke:2005hb}.
Depending on the quark model, in addition to LIRs, also further relations may 
arise~\cite{Signal:1996ct,Jakob:1997wg,Pasquini:2008ax,Avakian:2010br}
due to (spherical, spin-flavor) symmetries of model 
wave-functions~\cite{Lorce:2011zta}.

When we focus on the case of an unpolarized target within a quark model, the 
general decomposition of the correlator is completely specified by 3 terms,
\be\label{Eq:correlator-model}
   \Phi^q(P,p,S;\mbox{path}) 
   = M_N \, A^q_1 + \slashed{P}\,A^q_2 + \slashed{p}\,A^q_3 + \cdots,
\ee
where the dots denote $\mathsf{T}$-odd or polarization-dependent terms, 
or gauge-link related $B^q_i$ amplitudes. If we denote by $\la P|\cdots|P\ra$ 
the target spin-averaged matrix element, then the complete set of 
unpolarized $\mathsf{T}$-even TMDs is given by 4
TMDs, the twist-2 $f_1^q(x,p_T)$, the twist-3 $e^q(x,p_T)$ and 
$f^{\perp q}(x,p_T)$, and the twist-4 $f_4^q(x,p_T)$:
\begin{align}\label{Eq:correlator-TMDs1}
   \frac12\int\di p^- \,{\rm tr}[\Phi^q\gamma^+] = 
   \int \frac{\di z^-\di^2z_T}{2(2\pi)^3} \, 
   e^{i p \cdot z} \, \la P  |  \overline{\psi}(0)\gamma^+
   \psi(z)  |  P\ra|_{z^+=0} &= f_1^q(x,p_T),\\
   \label{Eq:correlator-TMDs2}
   \frac12\int\di p^- \,{\rm tr}[\Phi^q\mathds{1}] = 
   \int \frac{\di z^-\di^2z_T}{2(2\pi)^3} \, 
   e^{i p \cdot z} \, \la P  |  \overline{\psi}(0)\mathds{1}
   \psi(z)  |  P\ra|_{z^+=0}& = \frac{M_N}{P^+}\,e^q(x,p_T),\\
   \label{Eq:correlator-TMDs3}
   \frac12\int\di p^- \,{\rm tr}[\Phi^q\gamma_T^j] = 
   \int \frac{\di z^-\di^2z_T}{2(2\pi)^3} \, 
   e^{i p \cdot z} \, \la P |  \overline{\psi}(0)\gamma_T^j
   \psi(z)  |  P\ra|_{z^+=0} &= \frac{p_T^j}{P^+}\,f^{\perp q}(x,p_T),\\
   \label{Eq:correlator-TMDs4}
   \frac12\int\di p^- \,{\rm tr}[\Phi^q\gamma^-] = 
   \int \frac{\di z^-\di^2z_T}{2(2\pi)^3} \, 
   e^{i p \cdot z} \, \la P  |  \overline{\psi}(0)\gamma^-
   \psi(z)  |  P\ra|_{z^+=0} 
   &= \biggl(\frac{M_N}{P^+}\biggr)^{\!2}f_4^q(x,p_T).
\end{align}
In terms of the Lorentz-scalar amplitudes $A^q_i$, these unpolarized TMDs read in quark models
\begin{align}
   f^q_1(x,p_T)   &= 2P^+ \int\di p^- \left(A^q_2+xA^q_3\right),\label{Eq:TMD1}\\
   e^q(x,p_T)     &= 2P^+ \int\di p^- \, A^q_1, \label{Eq:TMD2} \\
   f^{\perp q}(x,p_T)&= 2P^+ \int\di p^- \, A^q_3,  \label{Eq:TMD3} \\
   f^q_4(x,p_T)   &= 2\left(\frac{P^+}{M_N}\right)^2
     \int\di p^- \left(P^-A^q_2+p^-A^q_3\right). \label{Eq:TMD4}
\end{align}
Up to twist-3 level in the unpolarized $\mathsf{T}$-even sector, we have
3 TMDs and 3 amplitudes. Thus, even in quark models, there are in general 
no relations between $f_1^q(x,p_T)$, $e^q(x,p_T)$ and $f^{\perp q}(x,p_T)$. 

The full structure of the quark correlator~\eqref{Eq:correlator} in the 
unpolarized $\mathsf{T}$-even sector is completed by the twist-4 TMD 
$f_4^q(x,p_T)$~\cite{Jaffe:1996zw,Goeke:2005hb}. Twist-4 TMDs are 
rather academic objects. In physical situations, like power 
corrections to the DIS structure functions, $f_4^q(x,p_T)$ mixes with 
other twist-4 quark-gluon correlators~\cite{Shuryak:1981dg,Jaffe:1981td,
Ellis:1982cd,Jaffe:1983hp,Qiu:1988dn,Ji:1993ey,Maul:1996dx,Geyer:2000ma}. 
While the practical understanding of power corrections is of interest~\cite{Alekhin:1999iq,
Blumlein:2008kz},
our motivation to include $f_4^q(x)$ is rather that it will serve as an 
important internal consistency check of our approach. In fact, in 
quark models new features emerge as one goes to higher twists 
(as in QCD, albeit on a far simpler level).

Including twist-4, we encounter in quark models the situation that 4 
unpolarized TMDs $\{f^q_1,\,e^q,\,f^{\perp q},\,f^q_4\}$ are expressed in terms 
of 3 amplitudes  $\{A^q_1,\,A^q_2,\,A^q_3\}$ (in QCD  the amplitude 
$B^q_1$ also contributes). This implies a LIR among these TMDs valid in Lorentz-covariant quark models (but not in QCD). Using the methods 
of~\cite{Tangerman:1994bb}, see App.~\ref{App:LIR}, we find
\be
    f_4^q(x) = \tfrac{1}{2}\,f_1^q(x)+\tfrac{{\ud}}{{\ud}x}f^{\perp q(1)}(x),
    \label{Eq:LIR-f4}
\ee
where $f_i^q(x)=\int\di^2p_T\,f_i^q(x,p_T)$ with $i=1,4$ and 
$f^{\perp q(1)}(x)=\int\di^2p_T\,\tfrac{p^2_T}{2M_N^2}\,f^{\perp q}(x,p_T)$.
To the best of our knowledge, this relation has not been presented in 
the literature before. Let us end this section with two general results.
In complete analogy to the positivity proof of $f_1^q(x,p_T)$, one
can show that the twist-4 TMD satisfies the positivity constraint
\be
    f_4^q(x,p_T) \ge 0. \label{Eq:f4-inequality} 
\ee
With $N_q$ denoting the valence quark number of flavor $q$, the following
sum rule is formally satisfied
\be
    2\int\di x\,f_4^q(x) = N_q. \label{Eq:f4-sum-rule}
\ee
We discuss in App.~\ref{App:f4-sum-rule} how this sum rule can be proven, 
and what is formal about it.

\section{Equations of motion relations in free quark models}
\label{Sec-3:eom-general-strategy}

Generally speaking, matrix elements of higher-twist operators can be 
decomposed by means of EOM into contributions 
from twist-2, mass terms and 
tilde-terms~\cite{Shuryak:1981dg,Jaffe:1981td,Ellis:1982cd}.
We present here a general approach to derive such relations tailored
for applications in quark models, where the situation is simplified
due to the absence of gauge interactions.
More precisely, in this section we concentrate on free quark models.
It should be noted that, for instance, parton model frameworks~\cite{Tangerman:1994eh,Jackson:1989ph,Efremov:2009ze,D'Alesio:2009kv,
Bourrely:2010ng,Anselmino:2011ch} belong to this class of models. 
After discussing the LFCQM in the next section, we 
will further generalize the formalism to models with interactions.

In order to derive a starting formula for EOM relations, we proceed as 
follows. Let $\Gamma$ be an arbitrary Dirac matrix. We apply the free
EOM within the fully unintegrated correlator, integrate by parts, and
obtain 
\begin{align}
0 &= \int \frac{\di^4z}{(2\pi)^4} \, 
      e^{i p \cdot z} \, \la P  |  \overline{\psi}(0)\Gamma
      (i\slashed{\partial} -m_q)\psi(z)  |  P\ra   \nonumber \\
  &= \int \frac{\di^4z}{(2\pi)^4} \, 
      (-i\partial_\mu e^{i p \cdot z}) \, \la P  |  \overline{\psi}(0)\Gamma 
      \gamma^\mu\psi(z)  |  P\ra   
   -  m_q\int \frac{\di^4z}{(2\pi)^4} \, 
      e^{i p \cdot z} \, \la P  |  \overline{\psi}(0)\Gamma
      \psi(z)  |  P\ra   \nonumber \\
  &= \int \frac{\di^4z}{(2\pi)^4} \, 
      e^{i p \cdot z} \, \la P | \overline{\psi}(0)\Gamma
      (\slashed{p} -m_q)\psi(z)  |  P\ra .
      \label{Eq:identity-IIa}
\end{align}
Next, repeating the above steps with $\la P |\overline{\psi}(-z)
(-i\overleftarrow{\slashed{\partial}}-m_q)\overline{\Gamma}\psi(0)|P\ra$ 
(or, equivalently, taking the complex conjugate 
of \eqref{Eq:identity-IIa}) and shifting the field positions by $z$, 
yields an identity analogous to~\eqref{Eq:identity-IIa} but with 
$\Gamma(\slashed{p} -m_q)$ replaced by $(\slashed{p} -m_q)\,\overline\Gamma$,
where $\overline\Gamma=\gamma^0\Gamma^\dag\gamma^0$ is the Dirac conjugate 
of $\Gamma$. Adding up these two identities yields
\be
0 =   \int\di p^-\int \frac{\di^4z}{2(2\pi)^4} \, 
      e^{i p \cdot z} \, \la P  |  \overline{\psi}(0)\left[
      (\slashed{p} -m_q)\,\overline{\Gamma} + \Gamma(\slashed{p} -m_q)
      \right]\psi(z)  |  P\ra,
      \label{Eq:identity-IIc}
\ee
where we introduced the $p^-$-integration and a factor $\tfrac{1}{2}$ 
for later convenience. In the following we also set $p^+=xP^+$.

Equipped with the identity~\eqref{Eq:identity-IIc}, we proceed to derive 
the EOM relations among $e^q(x,p_T)$, $f^{\perp q}(x,p_T)$ and $f^q_1(x,p_T)$.
They are obtained by choosing appropriate $\Gamma$ matrices. 
Choosing respectively $\Gamma=\gamma^+$ and $\Gamma=i\sigma^{+j}_T$,
we obtain 
\ba
     x\,e^q(x,p_T) &=& \frac{m_q}{M_N}\,f_1^q(x,p_T), \label{Eq:EOM-e-f1}\\
     x\,f^{\perp q}(x,p_T) &=& f_1^q(x,p_T), \label{Eq:EOM-f1-fperp}
\ea
which coincide with the EOM relations in QCD~\cite{Bacchetta:2006tn} 
but with (in free quark models) consistently neglected tilde-terms.

We remark that $\slashed{p}=\gamma^+p^-+\gamma^-p^+-\gamma_T^j p_T^j$ in the 
identity~\eqref{Eq:identity-IIc} introduces the factors $p^+=xP^+$ or $p_T^j$ which  
become prefactors of $x$ in~\eqref{Eq:EOM-e-f1} and~\eqref{Eq:EOM-f1-fperp}
or are ``absorbed'' by the definition of~\eqref{Eq:correlator-TMDs3}.
But the piece with $\gamma^+p^-$ drops out due to $(\gamma^+)^2=0$.
However, at twist-4 the component $p^-$ contributes. 
In order to eliminate it, we derive the identity
\begin{align}
   \label{Eq:trivial-and-enlightning-identity}
   0&=\int\frac{\di^4z}{(2\pi)^4}\, 
   e^{i p \cdot z} \, \la P  | \overline{\psi}(0)\Gamma 
   (i\slashed{\partial}+m_q)(i\slashed{\partial}-m_q)\psi(z) | P\ra \nonumber\\
&=\int\frac{\di^4z}{(2\pi)^4}\, 
   e^{i p \cdot z} \, \la P  | \overline{\psi}(0)\Gamma 
   (2xP^+p^--p_T^2-m_q^2)\psi(z)  |  P\ra ,
\end{align}
which reflects the fact that the correlator~\eqref{Eq:correlator} 
describes on-shell quarks in a free quark model. 
Using the identity~\eqref{Eq:identity-IIc} with $\Gamma=i\sigma^{+-}$ and 
making use of~\eqref{Eq:trivial-and-enlightning-identity}, we derive the 
EOM relation 
\be\label{Eq:EOM-f4-f1}
   x^2\,f^q_4(x,p_T)=\frac{p^2_T+m^2_q}{2M^2_N}\,f^q_1(x,p_T).
   \ee
For $\Gamma\in\{\mathds 1,\,\gamma^-,\,\gamma^j_T,
\,i\sigma^{-j}_T,\,i\sigma^{jk}_T\}$ we obtain linear 
combinations of the EOM relations~\eqref{Eq:EOM-e-f1},~\eqref{Eq:EOM-f1-fperp} and~\eqref{Eq:EOM-f4-f1}. 
 For example, for the choice $\Gamma=\mathds 1$ one obtains an 
 EOM connecting all 4 TMDs, which reduces to \eqref{Eq:EOM-f4-f1}
 using \eqref{Eq:EOM-e-f1} and \eqref{Eq:EOM-f1-fperp}, namely
 \be
 \label{Eq:EOM-unpol-all}
   x^2\,f^q_4(x,p_T) 
       = \frac{ p^2_T}{M^2_N}\,x\,f^{\perp q}(x,p_T)
       + \frac{m_q}{M_N}\,x\,e^q(x,p_T)
       - \frac{ p^2_T+m^2_q}{2M^2_N}\,f^q_1(x,p_T).
 \ee
All the other $\Gamma$-structures are not relevant 
for unpolarized TMDs. 

We end this section with three important remarks. First, in free quark models the set of unpolarized $\mathsf{T}$-even 
TMDs $\{f_1^q,\,e^q,\,f^{\perp q},\,f_4^q\}$ can be expressed in terms of 
one single TMD, say $f_1^q$. That there is only one independent 
structure, can also be seen as follows. In Eq.~(\ref{Eq:identity-IIa}) 
we have shown that in the class of free quark models 
${\rm tr}[\Gamma(\slashed{p}-m_q)\Phi^q]=0$ for all $\Gamma$.
This implies that $(\slashed{p}-m_q)\Phi^q=0$, and inserting here 
the decomposition~\eqref{Eq:correlator} of the correlator,
for the case of an unpolarized nucleon in a quark model, yields
\be\label{Eq:XXX}
0=(\slashed{p}-m_q)\,\Phi^q= (\slashed{p}-m_q) 
  (M_N \, A^q_1 + \slashed{P}\,A^q_2 + \slashed{p}\,A^q_3)
 =(\slashed{p}-m_q)(M_N\,A^q_1-m_q\,A^q_3) +
 (\slashed{p}-m_q) \slashed{P}\,A^q_2,
\ee
where we used $(\slashed{p}-m_q) \slashed{p} A_3 = -m_q(\slashed{p}-m_q)A_3$
if $p^2=m^2_q$. Since the Dirac matrices $\mathds 1$, $\slashed{p}$, 
$\slashed{P}$ and $\slashed{p}\slashed{P}$ are linearly independent 
for $p\not\propto P$, we conclude that
\be\label{Eq:amplitudes-free-EOM}
A^q_1=\frac{m_q}{M_N}\,A^q_3, \qquad A^q_2=0.
\ee
Using this result in Eqs.~\eqref{Eq:TMD1}-\eqref{Eq:TMD4} 
together with $2xP^+p^-=p^2_T+m^2_q$, we recover the 
relations~\eqref{Eq:EOM-e-f1},~\eqref{Eq:EOM-f1-fperp}, 
and~\eqref{Eq:EOM-f4-f1}. 
In particular, Eq.~\eqref{Eq:amplitudes-free-EOM} shows that in free quark models the unpolarized correlator consists of only one independent 
amplitude, meaning that all unpolarized TMDs are related to each other.

Second, since the general Lorentz decomposition in models
with on-shell quarks is fully specified by a single $A_i$ amplitude
according to~\eqref{Eq:amplitudes-free-EOM}, all our 
free EOM relations~\eqref{Eq:EOM-e-f1},~\eqref{Eq:EOM-f1-fperp} 
and~\eqref{Eq:EOM-f4-f1} can in some sense be understood as LIRs.
It has to be stressed, that the general LIR~\eqref{Eq:LIR-f4}
only explores Lorentz invariance in relativistic quark models, but 
makes no use of model details such as EOMs. Therefore, none of the EOM 
relations~\eqref{Eq:EOM-e-f1},~\eqref{Eq:EOM-f1-fperp},~\eqref{Eq:EOM-f4-f1}
is equivalent to the general LIR \eqref{Eq:LIR-f4}.
However, a particular linear combination of~\eqref{Eq:EOM-e-f1},~\eqref{Eq:EOM-f1-fperp} and~\eqref{Eq:EOM-f4-f1} can be {\it formally} proven 
to be equivalent to the LIR~\eqref{Eq:LIR-f4}. The proof is formal though, 
since it can be invalidated by the properties of the amplitudes $A_i^q$ 
in a given model, see App.~\ref{App:LIR-on-shell-case}.

Third, the EOM relations in interacting quark models
can be anticipated from Eqs.~\eqref{Eq:EOM-e-f1},~\eqref{Eq:EOM-f1-fperp} and~\eqref{Eq:EOM-f4-f1}, and read
\ba
     x\,e^q(x,p_T) &=& \;\;\,x\,\widetilde{e}^{\,q}(x,p_T) 
        \;+\; \frac{m_q}{M_N}\,f_1^q(x,p_T), \label{Eq:EOM-e-f1-QCD}\\
     x\,f^{\perp q}(x,p_T) &=& x\,\widetilde{f}^{\,\perp q}(x,p_T) 
        \;+\; f_1^q(x,p_T), \label{Eq:EOM-f1-fperp-QCD}\\
     x^2f^q_4(x,p_T) &=& \;\, x^2\widetilde{f}_4^{\,q}(x,p_T) 
        \;+\; \frac{p^2_T+m^2_q}{2M^2_N}\,f^q_1(x,p_T),\label{Eq:EOM-f4-f1-QCD}
 \ea
where the operator definitions of the specific tilde-terms have to be 
carefully worked out using the EOMs of the models under consideration. 
In QCD~\eqref{Eq:EOM-e-f1-QCD} and~\eqref{Eq:EOM-f1-fperp-QCD} hold, 
with the tilde-terms  defined in terms of quark-gluon 
correlators~\cite{Bacchetta:2006tn}.
But the term proportional to $p_T^2\,f_1^q(x,p_T)$ in~\eqref{Eq:EOM-f4-f1-QCD} could in QCD be naturally expressed in terms
of correlators with transverse gluon inclusions of the type 
$\la N|\bar\psi \,i\slashed{D}_T\gamma^+i\slashed{D}_T\psi|N\ra$~\cite{Ji:1993ey}. Our free quark model results are recovered in the 
limit $i\slashed{D}_T\to i\slashed{\partial}_T$. For QCD treatments
of higher-twist distributions we refer to~\cite{Shuryak:1981dg,Jaffe:1981td,
Ellis:1982cd,Jaffe:1983hp,Qiu:1988dn,Ji:1993ey,Maul:1996dx,Geyer:2000ma}.
We also remark that the ``brute-force'' systematic neglect of 
all QCD quark-gluon correlations is the basis for WW-type 
approximations~\cite{Avakian:2007mv}, and the general helicity formalism 
with the twist-2 QCD parton model of Ref.~\cite{Anselmino:2011ch}.

After discussing models where the quarks obey the free EOM 
(which is not necessarily the same as models without interactions) in 
Secs.~\ref{section-51} and~\ref{section-52}, we will come 
back to several interacting quark models in 
Sec.~\ref{Sec-7:eom-in-interacting-models}. 

\section{Ensemble of free quarks }
\label{section-51}

In this section we derive the general expression for the unpolarized T-even 
TMDs up to twist-4 in quark models in which the quarks obey the free Dirac 
equations. Following Ref.~\cite{Tangerman:1994eh}, we assume that
the nucleon is described as an ensemble of non-interacting partons of 
momentum $P$ and spin S, which can be considered as a generic 
prototype for parton-model approaches~\cite{Jackson:1989ph,Efremov:2009ze,
D'Alesio:2009kv,Bourrely:2010ng,Anselmino:2011ch}.
We consider the TMD correlator
\begin{equation}\label{correlator:unp}
  \Phi^{[\Gamma]q}(x, {\bm p}_T) \equiv
  \frac12\int\di p^-\,{\rm tr}[\Phi^q\Gamma] = 
  \int \frac{\di z^-\di^2z_T}{2(2\pi)^3} \, 
   e^{i p \cdot z} \, \la P  |  \overline{\psi}(0)\Gamma
   \psi(z)  |  P\ra|_{z^+=0},
   \end{equation}
where $\Gamma=\{ \gamma^+,\, \mathds{1}, \,\gamma_T^j,\, \gamma^-\}$
stands for the matrices entering the definition of unpolarized 
$\mathsf{T}$-even TMDs. In Eq.~\eqref{correlator:unp}, we insert the 
free-field Fourier expansion of the quark field $\psi$ on the surface 
$z^+=0$. We could equivalently use light-front as well as
instant-form quantization for free fields. However, to make the link 
with the LFCQM which will be discussed in the following section,  
we adopt the light-front form with the following Fourier expansion
\begin{equation}\label{fourier}
   \psi(z^-, \bm{z}_T)=\int\frac{{\rm d}k^+{\rm d}^2k_T}{2k^+(2\pi)^3}\,
   \Theta(k^+) \sum_\lambda\left\{
   b^q_\lambda(\tilde k)u_\lambda(\tilde k)\, e^{-ik^+z^-+i\boldsymbol{k}_T\cdot\boldsymbol{ z}_T}
   +d^{q\dagger}_\lambda(\tilde k)v_\lambda(\tilde k)\, 
   e^{ik^+z^- - i\boldsymbol{k}_T\cdot\boldsymbol{ z}_T}\right\},
   \end{equation}
where $b^q$ and $d^{q\dagger}$ are the annihilation operator of the quark field
and the creation operator of the antiquark field, respectively. 
Furthermore, $\lambda$ is the light-front helicity of the partons and $\tilde k$ denotes the light-front momentum variable $\tilde k=(k^+,\boldsymbol{k}_{T})$.
Using~\eqref{fourier} and restricting ourselves to the 
quark contribution, the operator in the correlator~\eqref{correlator:unp} reads
\begin{eqnarray}\label{operator}
   \int \frac{\di z^-\di^2z_T}{2(2\pi)^3} \, 
   e^{i p \cdot z} \, \overline{\psi}(0)\Gamma\psi(z) |_{z^+=0}
   &=& \frac{1}{2}\int\frac{{\rm d}k^+{\rm d}^2k_T}{2k^+(2\pi)^3}\,
   \Theta( k^+) \int\frac{{\rm d}k'^+{\rm d}^2k'_T}{2k'^+(2\pi)^3}\,\Theta( k'^+)
   \,\delta(p^+- k^+)\,
   \delta^{(2)}(\boldsymbol{p}_T-\boldsymbol{k}_T) \nonumber\\
   &&\times\sum_{\lambda,\lambda'}\overline u_{\lambda'}(\tilde{k}')\Gamma 
   u_\lambda(\tilde{k})\, b_{\lambda'}^{q\dagger}(\tilde{k}')b^{q}_{\lambda}(\tilde{k}).
   \end{eqnarray} 
 By inserting~\eqref{operator} in the correlator~\eqref{correlator:unp}, 
we obtain
\begin{eqnarray}\label{phi-overlap3}
   \Phi^{[\Gamma]q}(x,\boldsymbol{p}_T)&=&
   \sum_{\lambda,\lambda'}
   \frac{ \overline u_{\lambda'}(\tilde{p})\Gamma u_\lambda(\tilde{p})}{2p^+}
   \,{\cal P}^q_{\lambda\lambda'}(\tilde p),
   \end{eqnarray}
where $x=p^+/P^+$ and
\begin{eqnarray}
   {\cal P}^q_{\lambda\lambda'}(\tilde p)=\frac{1}{2(2\pi)^3}
   \int\frac{{\rm d}k'^+{\rm d}^2k'_T}{2k'^+(2\pi)^3}\,\Theta(k'^+)
   \langle P|b_{\lambda'}^{q\dagger}(\tilde{k}')b^q_\lambda(\tilde{p})|P\rangle.     
   \label{mom-density1}
   \end{eqnarray}
${\cal P}^q_{\lambda\lambda'}$ is a density matrix in the space of the quark 
light-front helicity and its trace 
\begin{eqnarray}
   {\cal P}^q(\tilde p)=\sum_{\lambda} {\cal P}^q_{\lambda\lambda}(\tilde p)
   \label{mom-density2}
   \end{eqnarray}
is the quark density operator evaluated in the target.
The light-front spinors are given by
\be
   u_+(\tilde p)=\frac{1}{\sqrt{2^{3/2}p^+}}
   \begin{pmatrix}
     \sqrt{2}\,p^++m_q\\\sqrt{2}\,p^R\\\sqrt{2}\,p^+-m_q\\\sqrt{2}\,p^R
   \end{pmatrix}, \qquad 
   u_{-}(\tilde p)=\frac{1}{\sqrt{2^{3/2}p^+}}
   \begin{pmatrix}
     -\sqrt{2}\,p^L\\\sqrt{2}\,p^++m_q\\\sqrt{2}\,p^L\\-\sqrt{2}\,p^++m_q
   \end{pmatrix},
   \ee
with $p^{R,L}=\tfrac{1}{\sqrt{2}}(p_x\pm i p_y)$. Specifying the matrix 
$\Gamma$ for the different unpolarized T-even TMDs, we find
\be\label{bilinears}
   \begin{aligned}
   \overline u_{\lambda'}(\tilde p)\gamma^+u_\lambda(\tilde p)
   =2p^+\,\delta_{\lambda\lambda'},\qquad\quad\,
   \overline u_{\lambda'}(\tilde p)u_\lambda(\tilde p)&=2m_q \,\delta_{\lambda\lambda'},\\
   \overline u_{\lambda'}(\tilde p)\gamma_{T}^ju_\lambda(\tilde p)=2p_{T}^j\,
   \delta_{\lambda\lambda'},\qquad
   \overline u_{\lambda'}(\tilde p)\gamma^-u_\lambda(\tilde p)
   &=\frac{p_T^2+m_{q}^2}{p^+}\,\delta_{\lambda\lambda'}.
   \end{aligned}
   \ee

Using these results in the quark correlator~\eqref{phi-overlap3}, we obtain 
\begin{align}
f_1^q(x,p_T)&= {\cal P}^q(\tilde p),\label{f1:lfwf}\\
x\,e^q(x,p_T)&=\frac{m_q}{M_N}\, {\cal P}^q(\tilde p),
\label{e:lfwf}\\
x\,f^{\perp q}(x,p_T)&={\cal P}^q(\tilde p),\label{fperp:lfwf}\\
x^2\,f^q_{4}(x,p_T)&=\frac{p_T^2+m_{q}^2}{2M_N^2}\,{\cal P}^q(\tilde p)
.\label{f4:lfwf}
\end{align}
From the results in Eqs.~\eqref{f1:lfwf}-\eqref{f4:lfwf}, it is obvious 
that the EOM relations~\eqref{Eq:EOM-e-f1},~\eqref{Eq:EOM-f1-fperp} 
and~\eqref{Eq:EOM-f4-f1} are satisfied.
These relations are a consequence of the on-shell relation 
for the single-quark states. In order to explicitly evaluate integrated relations such as~\eqref{Eq:LIR-f4}, we need to specify the quark momentum density~\eqref{mom-density1} and therefore a model for the target state.
To this aim, we will use as an example the LFCQM.

\section{Light-front constituent quark model }
\label{section-52}

The LFCQM has been used successfully to describe many nucleon 
properties~\cite{Boffi:2002yy,Boffi:2003yj,Pasquini:2004gc,
Pasquini:2005dk,Pasquini:2006iv,Pasquini:2007xz,Pasquini:2007iz,
Boffi:2007yc,Pasquini:2009ki,Lorce:2011dv} including leading-twist 
TMDs~\cite{Pasquini:2008ax,Boffi:2009sh,Pasquini:2010af,Pasquini:2011tk,
Pasquini:2014ppa}.
Here we extend the analysis to unpolarized $\mathsf{T}$-even TMDs beyond 
leading twist, restricting ourselves to the three-quark (3Q) Fock 
sector.
The light-front Fock-space expansion of the 
nucleon state is performed in terms of free on-mass-shell parton states 
with the essential QCD bound-state information encoded in the light-front 
wave-function (LFWF). 
Restricting ourselves to the 3Q Fock sector, one therefore 
effectively deals with an ensemble of free quarks as described in 
Sec.~\ref{section-51}. 
The $\mathsf T$-even unpolarized TMDs can be therefore 
expressed as in Eqs.~\eqref{f1:lfwf}-\eqref{f4:lfwf} where, as we 
will show, the quark momentum density in the proton  ${\cal P}^q$ 
is given by  the overlap of LFWFs averaged over the light-front 
helicity of the quarks. We will apply the results obtained in this 
section to a specific model for the LFWFs~\cite{Schlumpf:1994bc}, 
discuss numerical results, and compare to other models in 
Sec.~\ref{Sec-X:numerical-results}
(after a dedicated discussion how those models describe higher 
twist TMDs in Sec.~\ref{Sec-7:eom-in-interacting-models}).

Restricting ourselves to the 3Q Fock sector, the target state with definite four-momentum $P=[P^+,\tfrac{M^2}{2P^+},\uzero_T]$ and light-front helicity $\Lambda$ can be written as follows
\begin{equation}\label{LFWF}
|P,\Lambda\rangle=\sum_{\{\lambda_i\}}\sum_{\{q_i\}}\int\left[ \frac{\di x}{\sqrt{x}}\right]_3\,[\di^2p_T]_3\,\psi^{\Lambda;q_1q_2q_3}_{\lambda_1\lambda_2\lambda_3}(r)\,|\{\lambda_i,q_i,\tilde p_i\}\rangle,
\end{equation}
where $\psi^{\Lambda;q_1q_2q_3}_{\lambda_1\lambda_2\lambda_3}$ is the 3Q  LFWF with $\lambda_i$  and $q_i$ referring to the 
light-front helicity and flavor of quark $i$, respectively, $r$ stands for $(r_1,r_2,r_3)$ with $r_i=(x_i M_{0} , \boldsymbol{p}_{Ti})$, 
and $M_0$ denotes the mass of the non-interacting 3Q state.
We note that the single particle states in \eqref{LFWF} are on-shell, 
\emph{i.e.} $p^{-}_{i}=(p_{Ti}^{2}+m_{q}^{2})/2p^{+}_{i}$, 
but the 3Q Fock state is off-shell since 
$\sum_i p^{-}_{i}\neq P^{-}$, where $P^-=\frac{M_{N}^{2}}{2P^+}$ is the 
minus component of the nucleon momentum. Furthermore, 
for the 3Q Fock state one also has $\sum_i p_{i}^{+}=P^+$, 
whereas the  LFWF depends  on the plus component of the momenta of 
the non-interacting system of three quarks, \emph{i.e.} 
$k^{+}_{i}=x_i M_0$, which is related to $p^{+}_{i}$ by a longitudinal 
light-front boost. The integration measures in Eq.~\eqref{LFWF} are defined as
\begin{equation}
\begin{split}
\left[ \frac{\di x}{\sqrt{x}}\right]_3&\equiv\left[\prod_{i=1}^3 \frac{\di x_i}{\sqrt{x_i}}\right]\delta\!\!\left(1-\sum_{i=1}^3x_i\right),\\
[\di^2p_T]_3&\equiv\left[\prod_{i=1}^3\frac{\di^2p_{Ti}}{2(2\pi)^3}\right]2(2\pi)^3\,\delta^{(2)}\!\!\left(\sum_{i=1}^3\up_{Ti}\right).
\end{split}
\end{equation}

The calculation of the $\mathsf T$-even unpolarized TMDs proceeds along the lines outlined in the previous section.
The explicit expression for the quark-momentum density is obtained by inserting the LFWF expansion of the proton~\eqref{LFWF} in Eqs.~\eqref{mom-density1} and~\eqref{mom-density2}, with the result
 \begin{eqnarray}\label{phi-overlap}
{\cal P}^q(\tilde{p})&=&\frac{1}{2}\sum_\Lambda\sum_{\{\lambda_i\}}\sum_{\{q_i\}}\sum_{\{\lambda'_i\}}\sum_{\{q'_i\}}  
  \int\left[ \frac{\di x}{\sqrt{x}}\right]_3\,[\di^2p_T]_3\,  \int\left[ \frac{\di x'}{\sqrt{x'}}\right]_3\,[\di^2p'_T]_3\,
  \psi^{*\,\Lambda;q'_1q'_2q'_3}_{\lambda'_1\lambda'_2\lambda'_3}(r')\,
  \psi^{\Lambda;q_1q_2q_3}_{\lambda_1\lambda_2\lambda_3}( r )\nonumber\\ 
  &&\times \frac{1}{2(2\pi)^3}\int\frac{{\rm d}k'^+{\rm d}^2k'_T}{2k'^+(2\pi)^3}\,\Theta(k'^+)\,\sum_{\lambda}\,\sum_{j=1}^3\langle \lambda'_j,\, q'_j,\, \tilde p'_j| b_{\lambda'}^{q\dagger}(\tilde{k}')b^q_\lambda(\tilde{p})| \lambda_j,\, q_j,\, \tilde p_j\rangle  
  \prod_{i\ne j}\langle \lambda'_i,\, q'_i,\, \tilde p'_i|  \lambda_i,\, q_i,\, \tilde p_i\rangle.
 \end{eqnarray}
The  matrix elements and scalar products in Eq.~\eqref{phi-overlap} read
      \begin{align}
      \langle \lambda'_j,\, q'_j,\, \tilde p'_j| b_{\lambda'}^{q\dagger}(\tilde{k}')b^q_\lambda(\tilde{p})| \lambda_j,\, q_j ,\,\tilde p_j\rangle  &=\delta_{qq_j}\delta_{qq'_j}\delta_{\lambda\lambda_j}\delta_{\lambda'\lambda'_j}\,2p^+(2\pi)^3\,\delta(p^+-p^+_j)\,\delta^{(2)}(\boldsymbol{p}_T-\boldsymbol{p}_{Tj})\nonumber\\
      &\phantom{=}\times 2k'^+(2\pi)^3\,\delta(k'^+-p'^+_j)\,\delta^{(2)}(\boldsymbol{k}'_T-\boldsymbol{p}'_{Tj}),\label{matrix1}\\
          \langle \lambda'_i,\, q'_i,\, \tilde p'_i|  \lambda_i,\, q_i,\, \tilde p_i\rangle&=\delta_{q_iq'_i}\delta_{\lambda_i\lambda'_i}\,2p^+_i(2\pi)^3\,\delta(p'^+_i-p^+_i)\,\delta^{(2)}(\boldsymbol{p}'_{Ti}-\boldsymbol{p}_{Ti}).\label{matrix2}
          \end{align}
Using~\eqref{matrix1} and~\eqref{matrix2}, and performing 
the integrations over $\tilde{k}'$  
and the quark momenta 
$\tilde p'_i$, Eq.~\eqref{phi-overlap} becomes
          \begin{equation}\label{phi-overlap2} 
       {\cal P}^q(\tilde{p})=\frac{1}{2}\sum_\Lambda\sum_{\{\lambda_i\}}\sum_{\{q_i\}}\sum_j \prod_{i\ne j} \sum_{\lambda}
        \delta_{qq_j}\delta_{\lambda\lambda_j}
        \int[ \di x ]_3\,[\di^2p_T]_3\, \Theta(p^+)\,  \delta(x-x_j)\,\delta^{(2)}(\boldsymbol{p}_T-\boldsymbol{p}_{T\, j})\,|\psi^{\Lambda;q_1q_2q_3}_{\lambda_1\lambda_2\lambda_3}(r)|^2.
            \end{equation} 
          
In the case of $SU(6)$-symmetric  LFWF, the contributions from all 
quarks $q_i$ with $i=1,2,3$ are equal. 
We can choose to label the active quark with $i=1$ and multiply by 
three the corresponding contribution. Then, the final results for 
the unpolarized TMD correlators with $SU(6)$-symmetric LFWF reads   
     \begin{eqnarray}\label{phi-overlap2B} 
       {\cal P}^q(\tilde p)&=&\sum_{\lambda,\lambda_2,\lambda_3} \sum_{q_2q_3}
         \int\ud[23]\,  | \psi^{+;qq_2q_3}_{\lambda \lambda_2\lambda_3}( r ) |^2,
            \end{eqnarray} 
where we used the notation
\begin{equation}
\ud[23]=[\di x]_3\,[\di^2p_T]_3\,3\,\Theta(x)\,\delta(x-x_1)\,\delta^{(2)}(\up_T-\up_{T1}).
\end{equation}
After discussing other quark models in the next section, we will produce 
numerical results from Eqs.~\eqref{f1:lfwf}-\eqref{f4:lfwf} with the quark 
momentum density~\eqref{phi-overlap2B} obtained from the LFWFs of 
Ref.~\cite{Schlumpf:1994bc}. Before proceeding with that, let us 
discuss a general result concerning the integrated LIR~\eqref{Eq:LIR-f4}.

In the LFCQM, the LIR~\eqref{Eq:LIR-f4} is not satisfied. 
This result is generic, and does not depend on the specific model for the LFWF. 
We checked that the LIR~\eqref{Eq:LIR-f4} is not supported 
neither using the LFWF of Ref.~\cite{Schlumpf:1994bc} nor those
of Ref.~\cite{Faccioli:1998aq}. Moreover, we also assured ourselves that 
the LIR~\eqref{Eq:LIR-f4} is also not valid in the light-front constituent 
model of the pion~\cite{forthcoming}, which demonstrates that this feature 
does not depend on whether one deals with a three-body light-front Fock state 
$|qqq\ra$ as in the case of the nucleon, or a two-body light-front Fock state 
$|\bar q q\ra$ as in the case of the pion.

From a technical point of view, the non-compliance with the 
LIR~\eqref{Eq:LIR-f4} can be understood as follows.
In the integration of 
the LFWFs the relation $\sum_i k^+_i= M_0 \neq M_N$ with the off-shell 
energy condition $ \sum_ip^-_{i}\neq P^-$ comes into play, and spoils 
the relation which would be naively expected for 
 non-interacting quarks. 
The reason is that LFWFs 
represent the overlaps of the interacting state with free multiparton 
Fock states $\psi_n=\langle n|\psi\rangle$ and contain the information 
about the interaction. In the LFCQM, we truncate the Fock space to the 
three-quark sector and use the free EOM to write down the bad components 
of the quark field in the TMD correlator. It is therefore not surprising 
that the free EOM relations are satisfied for the unintegrated TMDs, 
where we single out the free-motion of the individual active quark from 
the spectator quarks. On the contrary, in the integrated TMDs, we convolute the 
motion of  the ``free'' active quark  with the dynamics of the interacting 
3Q system, with a consequent violation of the LIR. 

In a light-front approach, such as the LFCQM, the violation of the LIR~\eqref{Eq:LIR-f4} is an expected feature, and reflects general issues 
of the light-front approach with sum rules of higher-twist parton 
distributions and with matrix elements of the minus component of 
the electromagnetic current. This has been elucidated from various 
perspectives~\cite{Burkardt:1991hu,Brodsky:1998hn}.
In order to explain this point, we first remark that in the LFCQM $f_4^q(x)$
vanishes in the limits $x\to 0$ and $x\to 1$ (as do all other parton 
distribution functions and TMDs). Because of that we can integrate~\eqref{Eq:LIR-f4} over $x$ and derive in this way the sum rule~\eqref{Eq:f4-sum-rule}, see App.~\ref{App:f4-sum-rule}. 
Thus, in the LFCQM the integral $\int\di x\,f_4^q(x)$ 
receives contributions from the region $x>0$ only. 
However, as shown in Ref.~\cite{Burkardt:1991hu} in 1+1-dimensional
QCD calculations, the sum rule~\eqref{Eq:f4-sum-rule} is satisfied
only if one takes into account a $\delta(x)$-contribution which 
originates from zero modes in the light-front quantization and
whose existence can also be established using dispersion relation 
techniques~\cite{Burkardt:1991hu}. More on $\delta(x)$-contributions to parton distribution functions 
can be found in App.~\ref{App:sum-rule-other-cases}.
The description of light-front zero modes is beyond the scope of the 
LFCQM, and it is therefore not surprising that this model does not 
satisfy the sum rule~\eqref{Eq:f4-sum-rule} and the LIR from which
this sum rule can be derived (within this model).

Alternatively one can explain the non-compliance of the LFCQM with the 
sum rule~\eqref{Eq:f4-sum-rule} by observing that it is related to the
matrix elements of the minus component of the electromagnetic current.
The latter is of course conserved in the light-front approach. 
But for that, one has to consider contributions from higher light-front 
Fock states~\cite{Brodsky:1998hn} which are not accounted for in the LFCQM. 
\section{Equations of motion relations in models with interactions}
\label{Sec-7:eom-in-interacting-models}

In this section, we discuss three models with interactions: bag model, 
spectator model, chiral quark-soliton model. We  focus on formal
aspects. Numerical results from some of these models will be
presented in the next section.

\subsection{Bag model}\label{Sect:7a}

In the MIT bag model, relativistic (in our case massless) quarks are confined 
due to imposed boundary conditions inside a spherical cavity of radius $R_0$ 
fixed by the nucleon mass according to 
$R_0 M_N=4\omega$~\cite{Chodos:1974je,Jaffe:1974nj,Celenza:1982uk}. 
Here $\omega\approx 2.04$ is the dimensionless ``frequency'' of the lowest 
eigenmode whose momentum space  wave-function is given by 
\begin{equation}
    \varphi_{m}(\vec{p})=i\sqrt{4\pi}N R_0^3
    \left (\begin{array}{r} t_0(p)\chi_m\\
     (\sigma^i\widehat{p}^{\, i})\,t_1(p)\chi_m
    \end{array} \right ),
    \label{wp}
    \end{equation}
where $\sigma^i$ ($\chi_m$) denote Pauli matrices (spinors) and 
$\widehat{p}^{\, i} = p^i/p$ with $p=|\vec{p}|$. The normalization factor 
$N$ and the functions $t_l$ (expressed in terms of spherical Bessel 
functions $j_l$ with $l=0,\,1$) are given by
\begin{equation}
    \label{Eq:t0-t1}
    t_l(p)=\int\limits_0^1\di u\, u^2 j_l(upR_0)j_l(u\omega) , \qquad
    N=\left(\frac{\omega^3}{2R_0^3(\omega-1)\sin^2\omega}\right)^{1/2}.
    \end{equation}
We introduce the convenient notation of~\cite{Avakian:2010br}
\be\label{Eq:notation}
    A=\frac{16\omega^4}{\pi^2(\omega -1)j_0^2(\omega)M_N^2},\qquad
    p_z=xM_N-\frac{\omega}{R_0},\qquad
    \widehat{M}_N=\frac{M_N}{p}.
\ee
In this notation, with $t_l\equiv t_l(p)$ in the following and with
the $SU(6)$ spin-flavor symmetry factors $N_u = 2$ and $N_d=1$, 
the results for the $\mathsf{T}$-even unpolarized TMDs read
\be\label{Eq:TMDs-bag}
\begin{aligned}
  f_1^q(x,p_T) &= N_q\, A
  \left[t_0^2 +2\widehat{p}_z\,t_0 t_1 +t_1^2 \right],&
  e^q(x,p_T) &= N_q\, A
  \left[t_0^2 -t_1^2 \right], \\
  f_4^q(x,p_T) &= N_q\, A\,\tfrac{1}{2}
  \left[t_0^2 -2\widehat{p}_z\,t_0 t_1  +t_1^2 \right], &
  f^{\perp q}(x,p_T) &= N_q\, A
  \left[2\widehat{M}_N\,t_0 t_1 \right]. 
\end{aligned}
\ee
The collinear function $e^q(x)$ was discussed in~\cite{Jaffe:1991ra},
while $e^q(x)$ and $f_4^q(x)$ were calculated in~\cite{Signal:1996ct}.
Except for $f^q_4(x,p_T)$, all these TMDs were discussed in detail in~\cite{Avakian:2010br}. 

We will not investigate analytically the EOM relations in this model,
and content ourselves with a qualitative discussion. 
The best example to explain the origin of tilde-terms in the bag model 
is $e^q(x,p_T)$ for which the general decomposition is given by
$x\,e^q(x,p_T)=x\,\tilde{e}^q(x,p_T)$ for massless quarks. The bag model
quarks obey the free Dirac equation {\it inside} the cavity, and we 
know that the absence of interactions implies vanishing tilde-terms.
Thus, the result for $e^q(x,p_T)$ in Eq.~\eqref{Eq:TMDs-bag} is a 
{\it boundary effect}~\cite{Jaffe:1991ra}. This is a physically 
appealing result: the bag boundary ``mimics'' confinement and 
hence gluonic effects. In this sense, it can be viewed as a 
(crude) model for quark-gluon correlations~\cite{Jaffe:1991ra}.
Note that the massless bag model quarks are off-shell, 
$p^2=2xM_N\,\frac{\omega}{R_0}-x^2M_N^2- p_T^2\neq 0$.

The TMDs $f_4^q$, $f_1^q$ and $f^{\perp q(1)}$ satisfy the LIR~\eqref{Eq:LIR-f4}.
This can be proven analytically by repeating step by step the proof of a different 
LIR from the Appendix of~\cite{Avakian:2010br}. In our case, also a simpler proof 
is possible. Exploring the fact that the integrand $2\widehat{M}_N\,t_0t_1$
is a spherically symmetric function of 
$\vec p^{\, 2} = p_T^2+p_z^2$ with $p_z=xM_N-\tfrac{\omega}{R_0}$, we obtain
\be\label{Eq:LIR-bag-proof}
     \frac{\di}{\di x} f^{\perp q(1)}(x)
     = N_q\,A\int\di^2p_T\,\frac{p_T^2}{2M_N^2}\,
     \frac{\di[2\widehat{M}_N\,t_0t_1]}{\di p_T^2}\,
     \frac{\di p_T^2}{\di\vec p^{\, 2}}
     \frac{\di\vec p^2}{\di x}\,    
     = N_q\,A \int\di^2p_T
     \left[-2\widehat{p}_z\,t_0t_1\right] ,
\ee
where the last step follows after integration by parts.
Combining this result with the expressions for $f^q_1$ and $f^q_4$
in Eq.~\eqref{Eq:TMDs-bag} proves the LIR~\eqref{Eq:LIR-f4}.

That $f_4^q(x)$ satisfies the sum rule~\eqref{Eq:f4-sum-rule} can be
shown in two ways.\footnote{
   \label{footnote-bag-model}
   Let us recall that the bag wave-functions give rise to unphysical
   antiquark distributions $f_1^{\bar q}(x)<0$  at variance 
   with positivity.
   The TMDs also receive non-vanishing (though numerically very small) 
   support from the regions $|x|\ge 1$.
   These unphysical contributions must be included to satisfy sum rules,
   \emph{i.e.} the integration is over the whole $x$-axis. These 
   and other caveats of this simplest bag model version can be improved, 
   see for example~\cite{Schreiber:1991tc}.}
One way is to integrate the model expressions~\eqref{Eq:TMDs-bag} over $p_T$
and $x$, with $\di x = \di p_z/M_N$ according to Eq.~\eqref{Eq:notation}. 
Hereby the odd terms $(\pm \widehat{p}_z\, t_0t_1)$ in the expressions 
for $f_4^q$ and $f_1^q$ in Eq.~\eqref{Eq:TMDs-bag} vanish, implying that 
the integrals $2\int\di x\,f_4^q(x) = \int\di x\,f_1^q(x) = N_q$ are 
equally normalized. Alternatively, knowing from direct computation that 
in the bag model $\tfrac{\di}{\di x}f^{\perp q(1)}(x)$ is a continuous 
function at $x=0$ (which in general does not need to be the case,
see App.~\ref{App:f4-sum-rule}), one can integrate the above-proven LIR 
\eqref{Eq:LIR-f4} to verify~\eqref{Eq:f4-sum-rule}.

\subsection{Spectator model}\label{Sect:7b}

In the spectator model, 
one treats the intermediate states that can be inserted in the definition 
of the correlator~\eqref{Eq:correlator} as
effective degrees of freedom with quantum numbers of diquarks 
and definite masses.
Adopting the model of Ref.~\cite{Jakob:1997wg} for the diquark spectator 
system, one can write, for the contribution $\Phi^{[\Gamma]}_D$ of the 
diquark-type $D$ to the quark correlator \eqref{Eq:correlator},
\be
   \Phi_{D}^{[\Gamma]}(x,\boldsymbol{p}_T)=
   \left.\frac{\uTr[\tilde\Phi_D\Gamma]}{4(1-x)P^+}
   \right|_{p^2=xM^2_N-\tfrac{p^2_T+xm^2_D}{1-x}}
\ee
with
\be
   \tilde\Phi_D=\frac{|g(p^2)|^2}{2(2\pi)^3}\,
   \frac{(\uslash p+m_q)(\,\uslash\! P+M_N)(1+a_D\gamma_5\uslash S)
   (\uslash p+m_q)}{(p^2-m_q^2)^2},
\ee
where $m_D$ is the diquark mass, $a_D$ is a spin factor taking the values $a_s=1$ (scalar diquark) and $a_a=-1/3$ (axial-vector diquark), and $g(p^2)$ is a form factor 
that takes into account in an effective way the composite structure 
of the nucleon and the diquark. This form factor is often assumed to 
be~\cite{Melnitchouk:1993nk}
\be
   g(p^2)=N\,\frac{p^2-m^2_q}{|p^2-\Lambda^2|^\alpha},
\ee
where $\Lambda$ is a cut-off parameter and $N$ is a normalization constant. 
This choice has the advantage of killing the pole of the quark propagator.

The respective diquark-contributions to the $\mathsf{T}$-even 
unpolarized TMDs read
\be\label{Eq:TMDs-diquark}
\begin{aligned}
   f_1(x,p_T) &= B\,\frac{(m_q+xM_N)^2+p^2_T}{1-x},\\
   e(x,p_T) &= B\,
   \frac{(1-x)(m_q+xM_N)(m_q+M_N)
     -m^2_D(x+\tfrac{m_q}{M_N})-(1+\tfrac{m_q}{M_N})p^2_T}{(1-x)^2}, \\
   f^\perp(x,p_T) &= B\,
   \frac{(1-x^2)M^2_N+2m_qM_N(1-x)-m^2_D-p^2_T}{(1-x)^2},\\
   f_4(x,p_T) &= B\,
   \frac{(1-x)\left[(m_q+M_N)^2-m^2_D\right]+\frac{p^2_T+m^2_D}{M^2_N}
     \left[\frac{p^2_T+m^2_D}{(1-x)}-2M_Nm_q-(1+x)M^2_N\right]}{2(1-x)^2}, 
\end{aligned}
\ee
where we introduced for convenience
\be
   B=\frac{N^2}{2(2\pi)^3}
   \left[\frac{1-x}{p^2_T+\lambda^2_D(x)}\right]^{2\alpha}\qquad\text{with}\qquad \lambda^2_D(x)=(1-x)\Lambda^2+xm^2_D-x(1-x)M^2_N.
\ee
The flavor dependence is provided by $SU(4)$ symmetry
\be
f^u_1=\tfrac{3}{2}\,f_1\big|_{D=s}+\tfrac{1}{2}\,f_1\big|_{D=a},\qquad f^d_1=f_1\big|_{D=a},
\ee
and similarly for the other TMDs. Except for $f^q_4(x,p_T)$, all these TMDs were already obtained 
in~\cite{Jakob:1997wg}. Remarkably, the tilde-terms are simply given by
\be
\begin{aligned}
   x\,\tilde e(x,p_T)
   &=B\,\frac{p^2-m_q^2}{1-x}\,\left(x+\frac{m_q}{M_N}\right),\\
   x\,\tilde f^{\perp}(x,p_T)
   &=B\,\frac{p^2-m_q^2}{1-x},\\
   x^2\,\tilde f_4(x,p_T)
   &=B\,\frac{p^2-m_q^2}{1-x}\,\frac{1}{2M^2}
   \left[\left[(m_q+xM_N)^2+p^2_T\right]+x(1-x)M^2_N-\frac{x(p^2_T+m^2_D)}{1-x}\right],
\end{aligned}
\ee
which illustrates the connection between interaction and quark off-shellness 
$p^2-m^2_q$. Using the analytic expressions~\eqref{Eq:TMDs-diquark}, it is 
straightforward to check that the LIR~\eqref{Eq:LIR-f4} is satisfied, 
see also App.~\ref{App-diquark} for further details.

\subsection{\boldmath Chiral quark-soliton model}

We proceed with the chiral quark-soliton model ($\chi$QSM). Here the nucleon is described as a chiral 
soliton in an effective, non-renormalizable low-energy theory ~\cite{Diakonov:1987ty} defined in terms of the Lagrangian 
${\cal L}=\overline\psi(i\slashed{\partial}-M\,U^{\gamma_5}-m_q)\psi$,
where $U^{\gamma_5}=\exp(i\gamma_5\vec\tau\cdot\vec\pi/f_{\pi})$
denotes the chiral field and $f_\pi = 93\,{\rm MeV}$ the pion decay constant.
The parameter $M=350\,{\rm MeV}$ is not a ``constituent quark mass'', 
but a dimensionful coupling constant of the quark fields to the chiral 
field, which is dynamically generated by instanton--anti-instanton 
interactions in the semi-classical description of the QCD vacuum~\cite{Diakonov:1983hh,Diakonov:1985eg,Diakonov:1995qy}.
A popular jargon is to refer to $M$ as ``dynamical mass''. 
In contrast, $m_q={\cal O}(\mbox{few MeV})$ is the current quark mass 
of light quarks. In many practical calculations, one can work in the 
chiral limit, and set $m_q$ to zero. In the following analytical 
derivations, we shall keep $m_q$ finite. The cutoff 
$\Lambda_{\rm cut}={\cal O}(\rho_{\rm av}^{-1})$ of the effective 
theory is set by the inverse of the average instanton size 
$\rho_{\rm av}^{-1}\approx 600\,{\rm MeV}$ and determines the 
initial scale of the model.
The theory can be solved in the limit of a large number of colors 
$N_c$, where a soliton solution is found for a static pion field with 
hedgehog symmetry $\vec\pi(\vec{x}) = f_\pi \vec e_r P(r)$ where 
$\vec{e}_r=\vec{x}/r$ and $r=|\vec{x}|$. Expressed in terms of the 
profile function $P(r)$, the chiral field is given by
$U^{\gamma_5}=\cos P(r)+i\gamma_5\,(\vec e_r\cdot\vec\tau)\sin P(r)$.

In the $\chi$QSM, the equation of motion is 
$(i\slashed{\partial}-M\,U^{\gamma_5}-m_q)\psi=0$. The difference
with the free quark case is the presence of the interaction term 
$M\,U^{\gamma_5}$, which will be responsible for the emergence of
``interaction-dependent'' tilde-terms. Since the interaction
contains no derivatives and $\overline{U^{\gamma_5}}=U^{\gamma_5}$, 
we obtain an identity analogue to Eq.~\eqref{Eq:identity-IIc},
with the substitution $m_q \mapsto m_q+M\,U^{\gamma_5}$, i.e.\
\be
0 = \int\di p^-\int \frac{\di^4z}{2(2\pi)^4} \, 
      e^{i p \cdot z} \, \la P  |  \overline{\psi}(0)\left[ (\slashed{p} -m_q-M\,U^{\gamma_5})
      \overline\Gamma+
      \Gamma(\slashed{p} -m_q-M\,U^{\gamma_5})\right]
      \psi(z)  |  P\ra.
      \label{Eq:identity-cQSM}
\ee
The treatment of the $(\slashed{p}-m_q)$ part is precisely the same as in 
Sec.~\ref{Sec-3:eom-general-strategy}, so we can focus on the structure 
$\left[\Gamma M\,U^{\gamma_5}+M\,U^{\gamma_5}\,\overline\Gamma\right]$.
Choosing respectively $\Gamma=\gamma^+$ and $\Gamma=i\sigma^{+j}_T$
yields
\ba
x\,e^q(x,p_T)&=&x\,\tilde{e}^q(x,p_T)+\frac{m_q}{M_N}\,f_1^q(x,p_T),
\label{Eq1cQSM}\\
x\,f^{\perp q}(x,p_T)&=& f_1^q(x,p_T),
\label{Eq3cQSM}
\ea
where 
\be
	x\,\tilde{e}^q(x,p_T)
	= \int \frac{\di z^-\di^2z_T}{2(2\pi)^3} \, 
   	  e^{i p \cdot z} \, \la P  |\overline{\psi}(0)\;
	  \frac{\{\gamma^+,MU^{\gamma_5}\}}{2M_N}
      	  \;\psi(z) |  P\ra\big|_{z^+=0}.
\ee
Several comments are in order. First, even in the chiral limit $m_q\to 0$, 
the $\chi$QSM predicts a non-zero $e^q(x,p_T)$ which arises from the 
interaction term. That the operator $MU^{\gamma_5}$ is associated 
with interactions is evident: it is proportional (a) to $M$ which is
the dynamically generated mass due to interactions of light quarks in 
the strongly interacting QCD (instanton) vacuum, and (b) to the chiral
field binding the effective quark degrees of freedom to form the
nucleon.
Second, it is remarkable that the strong chiral interactions 
do not generate a tilde-term in the case of $f^{\perp q}(x,p_T)$. 
This information is very useful for phenomenology. In fact, it supports 
the WW-approximation for this TMD, which was applied to phenomenology 
in~\cite{Anselmino:2005nn}. Third, the above expressions describe also 
antiquark TMDs according to
\be\label{Eq:C-parity}
\text{TMD}^{\bar q}(x,p_T)=\pm\text{TMD}^{q}(-x,p_T)
\ee
with a $(-)$-sign for $f_1$, and a $(+)$-sign for $e$ and $f^\perp$.

It is interesting to remark that, from the $\chi$QSM, we can recover 
results of the non-interacting theory by taking formally the limit 
$U\to \mathds 1$, which can be done by letting the size of the soliton 
go to zero~\cite{Diakonov:1987ty}. In this formal limit
\be\label{Eq:limit-non-interacting}
      \lim\limits_{{\rm non-}\atop{\rm interact.}} x\,\tilde e^q(x,p_T)= 
      \frac{M}{M_N}\,
      \lim\limits_{{\rm non-}\atop{\rm interact.}} f_1^q(x,p_T).
\ee
Interestingly the tilde-term does not vanish but becomes effectively
a mass term.\footnote{
   \label{Footnote-chiQSM}
   In $\chi$QSM calculations vacuum subtraction is
   understood, \emph{i.e.} symbolically $\la N|\,\dots\,|N\ra\equiv
    \la N|\,\dots\,|N\ra_{U^{\gamma_5}}-\la N|\,\dots\,|N\ra_{U^{\gamma_5}\to 1}$,
   showing that  in the formal limit $U^{\gamma_5}\to 1$ the contribution from 
   the Dirac continuum is cancelled in Eq.~\eqref{Eq:limit-non-interacting}.
   The contribution from the discrete level however remains~\cite{Diakonov:1987ty} and is responsible for the limits in 
   Eq.~\eqref{Eq:limit-non-interacting}.}
By taking $U^{\gamma_5}\to 1$, we ``removed'' the soliton field which 
binds the quarks. But we did not remove effects of the strongly 
interacting QCD vacuum where our quarks are  embedded.
In fact, light current quarks (with small masses $m_q$) acquire $M$ 
as a response to collective instanton vacuum effects. So $M$ and
hence the result in Eq.~\eqref{Eq:limit-non-interacting} are 
of dynamical origin. If we ``switched off'' QCD vacuum effects,
also $M\to0$. Thus, the result for $\widetilde{e}^{\, q}(x,p_T)$
is clearly an interaction-dependent tilde-term. 
It is an important cross check that, in the formal 
limit of vanishing interactions, we recover results from the free theory.

It should be noted that in QCD the first two Mellin moments of 
$\tilde{e}^q(x)$ vanish (see Eq.~\eqref{Eq:e-mas-tilde-sum-rules} 
in App.~\ref{App:f4-sum-rule}), but not in the $\chi$QSM. 
This is a limitation of the model, but not its failure. 
QCD sum rules for Mellin moments are specific to gauge 
theories (one would have basically the same sum rules in QED). 
The model interactions are different, which results in different but 
consistently satisfied sum rules within the models~\cite{Schweitzer:2003uy}.

For completeness, we also discuss $f_4^q(x,p_T)$. Exploring hedgehog 
symmetry, one finds that the sum rule~\eqref{Eq:f4-sum-rule} holds. 
Positivity can be proven within the model in complete analogy to
$f_1^q(x,p_T)$~\cite{Efremov:2002qh}. In order to derive the EOM relation, 
we use the identity~\eqref{Eq:identity-cQSM} with \emph{e.g.} 
$\Gamma=\mathds 1$. 
As in free quark case, one encounters a contribution from the structure 
$\gamma^+p^-$ where one has to eliminate $p^-$. This could be
done by generalizing~\eqref{Eq:trivial-and-enlightning-identity} to the case 
of the $\chi$QSM through the replacement of
$(i\slashed{\partial}+m_q)
 (i\slashed{\partial}-m_q)$
by 
$(i\slashed{\partial} +M\,U^{-\gamma_5}+m_q)
 (i\slashed{\partial} -M\,U^{ \gamma_5}-m_q)$
with due care to the fact that  $U^{\pm\gamma_5}$ do not need to commute 
with $\Gamma$. Alternatively, we can explore 
~\eqref{Eq:identity-cQSM} with $\Gamma=i\sigma^{+-}$ where the
structure $\gamma^+p^-$ appears in a different linear combination,
allowing us to eliminate it as a whole. After
straightforward algebra we obtain
\be
      x^2\,f_4^q(x,p_T) =
       x^2\,\tilde{f}_4^q(x,p_T)+\frac{{p}_T^2+m_q^2}{2M_N^2}\,f_1^q(x,p_T)
\ee
with (for the flavor-singlet case)
\be\label{Eq:f4-tilde-in-CQSM}
      x^2\,\tilde{f}_4^q(x,p_T) =   
      \int\frac{\di z^-\di^2z_T}{2(2\pi)^3}\, e^{i p \cdot z} \, 
      \la P|\overline{\psi}(0)\,\Biggl(\frac{m_q}{M_N}\;
	  \frac{\{\gamma^+,M\,U^{\gamma_5}\}}{2M_N} 	
	+  \frac{xP^+M U^{\gamma_5}}{M_N^2}\Biggr)
      \psi(z)  |  P\ra\big|_{z^+=0}, 
\ee
where all terms are matrix elements of the chiral field $U^{\gamma_5}$ 
and proportional to the dynamical mass $M$, \emph{i.e.} manifestly
interaction-dependent.\footnote{
   Eq.~(\ref{Eq:f4-tilde-in-CQSM}) was incorrect in the first archive 
   version arXiv:1411.2550v1 and in the journal version of this article.
   Our original conclusion, namely that $\tilde{f}_4^q(x,p_T)$ is manifestly
   interaction dependent, is not altered by this correction.}
Thus, switching off (soliton, instanton vacuum) 
interactions removes $\tilde{f}_4^q(x,p_T)$.

In the language of the light-front Fock-state expansion, the quark correlator 
in that model contains all $|nq,(n-3)\bar q\ra$ components for 
$n=3,\,4,\,5,\,\cdots$ summed up.
The calculation in terms of a Fock expansion is efficient if one
restricts oneself to the minimal Fock state $n=3$~\cite{Lorce:2011dv},
and becomes quickly impractical beyond that~\cite{Lorce:2006nq,Lorce:2007as,Lorce:2007fa}. To get the ``full answer'',
one has to evaluate the entire correlator. This numerically laborious 
task was done at large $N_c$ for the flavor-singlet unpolarized TMDs $f_1^{u+d}(x,p_T)$ 
and $f_1^{\bar u+\bar d}(x,p_T)$~\cite{Wakamatsu:2009fn,Schweitzer:2012hh},
from which we could immediately obtain flavor-singlet results for
$xf^{\perp q}(x,p_T)$ via~\eqref{Eq3cQSM}, but the computation in the 
non-singlet channel has not yet been performed. Results for the parton 
distribution function $e^q(x)$ were presented in~\cite{Schweitzer:2003uy,Wakamatsu:2003uu},
while $f_4^q(x)$ was never studied. 

In this section we treated the $\chi$QSM as a ``quark model'' as 
done in~\cite{Schweitzer:2003uy,Wakamatsu:2003uu} and other higher-twist studies~\cite{Wakamatsu:2000ex,Gamberg:1998vg}. 
As in any quark model, also here it is possible to evaluate matrix elements of 
$\bar\psi\Gamma\psi$ operators of any twist.
We found the results consistent in the sense that the tilde-terms, 
which we separated off by means of EOM, really encode model interactions 
and vanish in a formal limit of a non-interacting theory.
However, strictly speaking the $\chi$QSM should be understood as the 
``leading-order'' approximation of the instanton vacuum model, 
which is of paramount importance to identify the 
model distributions of quarks and antiquarks with {\it leading-twist} 
QCD parton distribution functions at low normalization scale 
$\mu_0\sim \rho^{-1}_{\rm av}\sim 600$ MeV~\cite{Diakonov:1996sr}. 
The tight connection of the $\chi$QSM to instanton vacuum became 
also apparent in our discussion: we were able to show that tilde-terms 
vanish, only after switching off all interactions, also those associated 
with instanton vacuum effects. 
Fully consistent higher-twist studies require to work directly in the 
instanton vacuum~\cite{Diakonov:1983hh,Diakonov:1985eg,Diakonov:1995qy}.
Only in this way a realistic description of the non-perturbative quark-gluon 
dynamics can be obtained. In some cases, tilde-terms of partons distribution
functions were found to be small~\cite{Balla:1997hf,Dressler:1999hc} in the 
instanton vacuum, but not in all~\cite{Diakonov:1995qy,Dressler:1999zi}.
Since a fully consistent treatment of higher-twist matrix elements 
requires instanton vacuum techniques, we refrain from showing here 
numerical results for higher-twist TMDs within the $\chi$QSM, and
refer to instanton vacuum model 
studies~\cite{Diakonov:1995qy,Balla:1997hf,Dressler:1999hc,Dressler:1999zi}.

\section{Numerical results}
\label{Sec-X:numerical-results}

In this section we present numerical results from the models.
The LFCQM results are new, and discussed in more detail. 
For comparison we include bag and spectator model results 
\cite{Jakob:1997wg,Avakian:2010br}.
All results refer to a low quark model scale. We then evolve $e^q(x)$ 
obtained in the LFCQM and compare it with a recent 
extraction~\cite{Courtoy:2014ixa}.

We apply the general light-front formalism elaborated in Sec.~\ref{section-52} 
to the model of 3Q LFWF from~\cite{Schlumpf:94a,Schlumpf:94b}.
The parameters of this model were fixed
to reproduce the anomalous magnetic moments of the proton and neutron.
The parameter of importance for the following discussion is the 
constituent quark mass $m_q=263$ MeV. The results of this
quark model, as well as any quark model without explicit gluon and
sea-quark degrees of freedom, refer to a low initial scale 
$\mu_{0\,{\rm LO}}=420$ MeV.
\begin{figure}[t!]
   \begin{center}
   \epsfig{file=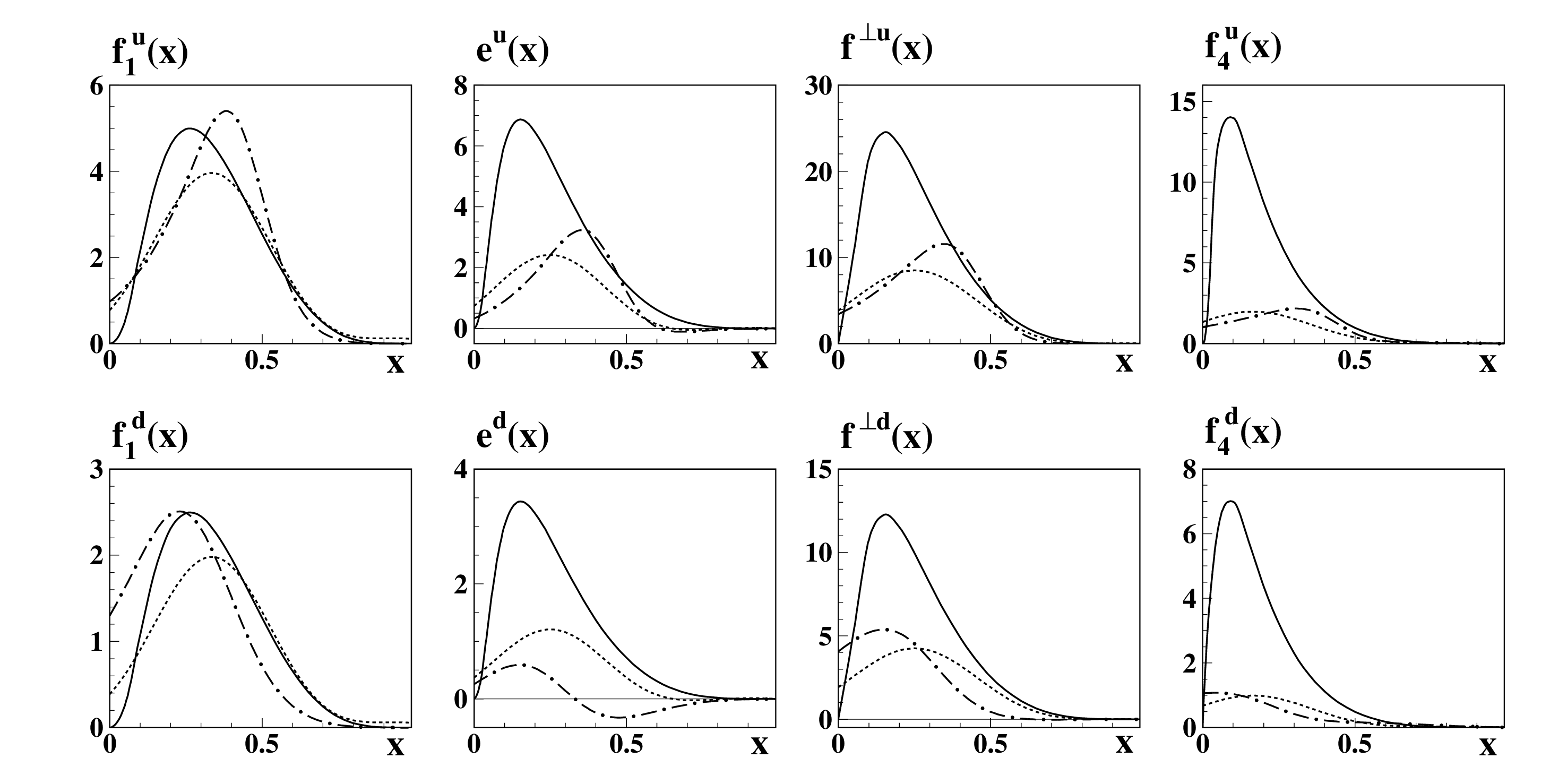,width=\columnwidth}
   \end{center}
   \vspace{-0.5 truecm}
   \caption{\footnotesize{
      $\mathsf{T}$-even spin-independent TMDs for up (upper panel) and 
      down (lower panel) quarks from different model calculations: 
      LFCQM (solid curves) obtained in this work;
      diquark model (dashed-dotted curves) of Ref.~\cite{Jakob:1997wg};
      bag model (dashed curves) of Ref.~\cite{Avakian:2010br}.}}
\label{fig1}
\end{figure}

The results for the integrated TMDs 
$f_1^q(x)$, $e^q(x)$, $f^{\perp q}(x)$, and $f_4^q(x)$ 
obtained from this approach are shown in Fig~\ref{fig1}.
The bag and spectator model results for $f_1^q(x)$, $e^q(x)$, and $f^{\perp q}(x)$
included for comparison in Fig.~\ref{fig1} are from~\cite{Avakian:2010br,Jakob:1997wg}, while the results for 
$f^q_4(x)$ in those models are from this work.
Because of $SU(6)$ spin-flavor symmetry, the flavor dependence of 
unpolarized $\mathsf{T}$-even TMDs is trivial in LFCQM and bag model, 
\emph{i.e.} $u$-quark distributions are a factor 2 bigger than
$d$-quark distributions. This is different in the diquark spectator 
model, where the two-body wave-function obeys $SU(4)$ symmetry that does 
not lead to a simple relation between $u$- and $d$-quark distributions 
(only in the large $N_c$-limit, where the scalar and axial diquark masses
become equal, would one have the same trivial flavor dependence in the
spectator model as in the other two models).

The results for the twist-2 function are comparable in the three models.
For instance, $f_1^q(x)$ exhibits a peak roughly around $x\approx 0.3$
in all models. Also the magnitude is similar, which is understandable 
because the flavor number sum rule determines the normalizations of 
the lowest moments.
In particular, the results from the bag model and LFCQM show a very 
similar behavior at large $x$. 

The picture is very different for higher twist.
As compared to the other models, the magnitude of 
the higher-twist TMDs is bigger in the LFCQM.
This is partly due to the fact that higher-twist TMDs in this model 
arise from mass effects, and the constituent quark mass of this model $m_q = 263$ MeV
is sizable.
Also the overall shapes of $e^q(x)$, $f^{\perp q}(x)$ and $f_4^q(x)$
differ largely in the three models. For instance, the maxima of the curves
are scattered over a wide interval in $x$. A very distinctive feature 
is the node in $e^d(x)$ in the diquark model. All models comply with the 
positivity constraint for $f_4^q(x)$ in Eq.~\eqref{Eq:f4-inequality}.

Note also that the distributions do not vanish at $x=0$ in the bag and 
diquark  models, in contrast to the LFQCM. This is due to the 
power-law ansatz of the model for the LFWF, which vanishes when 
any $x_i\rightarrow 0$. 
As a consequence\footnote{
    \label{footnote-LFCQM}
    This small-$x$ behavior is at variance with QCD and one of the reasons 
    why small $x$ are beyond the range of applicability of the LFCQM approach~\cite{Boffi:2009sh}. We will comment on this in more detail below.}
$f_1^q(x)\propto x^3$  for $x\to 0$. 
Therefore $e^q(x)$, $f^{\perp q}(x)$ and 
$f_4^q(x)$, which are related to $f_1^q(x)$ by means of the EOM 
relations~\eqref{Eq:EOM-e-f1},~\eqref{Eq:EOM-f1-fperp} and~\eqref{Eq:EOM-f4-f1}, 
have not only regular small-$x$ limits but even vanish for $x\to 0$, too.

Next let us discuss sum rules. The twist-3 parton distribution function
$e^q(x)$ obeys the sum rules~\cite{Jaffe:1991ra}
\ba\label{Eq:e-Mellin-mom-1}
\int\di x\,e^q(x) &=&\frac{1}{2 M_N}\,\la P|\overline\psi(0)\psi(0)|P\ra ,\\
\int\di x\,x\,e^q(x)&=&\frac{m_q}{M_N}\,N_q. \label{Eq:e-Mellin-mom-2}
\ea
The LFCQM satisfies both sum rules. In the case of~\eqref{Eq:e-Mellin-mom-1}
this means that integrating $e^q(x)$ yields the same result as evaluating
the local matrix element on the right-hand-side of this equation.
The compliance of the model with the second sum rule is evident from the 
EOM relation~\eqref{Eq:EOM-e-f1}. For the first sum  rule, however, this 
is highly non-trivial. We explain this in detail in 
App.~\ref{App:sum-rule-other-cases}.
Numerically the result for the sum rule~\eqref{Eq:e-Mellin-mom-1} is
2.22 (1.11) for up (down) quarks, and $\sum_q\int\di x\,e^q(x)=3.33$ 
at the low scale of the model. Evolving this result to a typical 
DIS scale of, say, $Q^2=1.5\,{\rm GeV}^2$ (see below), yields  
\be\label{Eq:e-Mellin-mom-1-num-evolved}
   \sum\limits_{q=u,d}\int\di x\,e^q(x)\biggl|_{Q^2=1.5\,{\rm GeV}^2} 
   =3.33 \times \frac{m_q(\mu_0^2)}{m_q(Q^2)} 
   = 8.29,
\ee
in agreement with $\sum_q\int\di x\,e^q(x) = $~(6--10) expected in QCD~\cite{Efremov:2002qh} (see App.~\ref{App:sum-rule-other-cases} for 
further comments on this result).

\begin{table}[t!]
\begin{tabular}{cc}

      \begin{tabular}{c|ccc}
      \hline
      \hline
      { }
      & \multicolumn{3}{c}{ \ LFCQM} \cr
      \hline
      TMD & \ $\la p_T\ra$ \ & \ $\la p_T^2\ra$ \ & \ $R_G$ \ \ \cr
      \hline
      $f_{1}^{q}$  & 0.24 & 0.080 & 0.96 \cr 
      $e^{q}$     & 0.22 & 0.069 & 0.95 \cr
      $f^{\perp q}$ & 0.22 & 0.069 & 0.95 \cr
      $f_{4}^{  q}$ & 0.28 & 0.110 & 0.95 \cr
      \hline
      \hline
      \end{tabular} \hspace{5mm} 
&      
      \begin{tabular}{c|c|c|c|c}
      \hline
      \hline
      { }
      & \ \ LFCQM \ \  
      & \ bag model \
      & spectator $(u)$ 
      & spectator $(d)$ \cr
      \hline
      TMD
      &  $\la p_{T,v}^2\ra^{1/2}$ 
      &  $\la p_{T,v}^2\ra^{1/2}$ 
      &  $\la p_{T,v}^2\ra^{1/2}$ 
      &  $\la p_{T,v}^2\ra^{1/2}$ \cr
      \hline
      $f_1^{q}$    & 0.135     & 0.28   & 0.20 & 0.27 \cr
      $e^q$       & 0.135     & 0.23   & 0.16 & 0.18 \cr
      $f^{\perp q}$ & 0.135   & 0.27   & 0.18 & 0.23 \cr
      $f_4^{q}$    & 0.200    & 0.17   & 0.18 & 0.25 \cr
      \hline
      \hline
      \end{tabular}
\cr   
      { }  & { } \cr
      {\bf\boldmath (a)} &  {\bf\boldmath (b)} 

\end{tabular}
\caption{\label{Table:pT-model}
   (a)
   $\la p_T\ra$ in units of GeV and $\la p_T^2\ra$ in units of GeV$^2$  
   as defined in Eq.~\eqref{Eq:define-mean-pT}, and the ratio 
   $R_G\equiv 2\la p_T\ra/(\pi\,\la p_T^2\ra)^{1/2}$ 
   for unpolarized $\mathsf T$-even TMDs from LFCQM.
   (b)
   The Gaussian widths $\la p_{T,v}^2\ra^{1/2}$ in units of GeV for unpolarized 
   $\mathsf T$-even TMDs in the valence-$x$ region at $x_v=0.3$ from LFCQM (here), 
   bag model~\cite{Avakian:2010br} and spectator model~\cite{Jakob:1997wg}.
   Using these widths in a Gaussian Ansatz, one approximates the true 
   $p_T$-dependence of the model TMDs within (5--20)$\%$ for 
   $p_T^2\lesssim 2\la p_{T,v}^2\ra$, see text.
   The spectator model results depend on the flavor $q=u,d$. 
   Results from LFCQM and bag model are 
   the same for $u$ and $d$.}
\end{table}

Next we turn our attention to the $p_T$-dependence of TMDs. We define the 
mean transverse momenta $(n=1)$ and the mean squared transverse momenta 
$(n=2)$ in the TMD as follows 
\be\label{Eq:define-mean-pT}
       \la p_{T}^n\ra= \frac{\int\di x\int\di^2p_T \,p_T^n\,\text{TMD}(x,p_T)}
                           {\int\di x\int\di^2p_T \,\text{TMD}(x,p_T)}.
\ee
In Table~\ref{Table:pT-model}$\,$(a) we show results for these quantities
for unpolarized $\mathsf{T}$-even TMDs in the LFCQM. Since in the 
LFCQM the flavor dependence appears as an overall factor $N_q$, 
the $\la p_T^n\ra$ in Eq.~\eqref{Eq:define-mean-pT} are equal
for $u$- and $d$-quarks. 
Compared to $f_1^q(x,p_T)$, the mean transverse momenta in $e^q(x,p_T)$ and 
$f^{\perp q}(x,p_T)$ are smaller while those of $f_4^q(x,p_T)$ are larger,
implying that $e^q(x,p_T)$ and $f^{\perp q}(x,p_T)$ fall off with $p_T$ 
faster than $f_1^q(x,p_T)$ and \emph{vice-versa} for $f_4^q(x,p_T)$.
An instructive quantity is the ratio 
$R_G\equiv 2\la p_T\ra/(\pi\,\la p_T^2\ra)^{1/2}$. If the $p_T$-dependence
of the TMDs was exactly Gaussian, this ratio would be unity.
Table~\ref{Table:pT-model}$\,$(a) shows that the LFCQM supports this
``measure of Gaussianity'' within $5\%$.

The definitions of $\la p_T^n\ra$ in Eq.~\eqref{Eq:define-mean-pT} 
are not useful in all models. 
In the bag model, the $x$-integration in~\eqref{Eq:define-mean-pT} 
would include unphysical regions and bias the result, 
see footnote~\ref{footnote-bag-model}. Moreover, $\la p_T^2\ra$  
defined in~\eqref{Eq:define-mean-pT} is divergent for some 
TMDs~\cite{Avakian:2010br}. 
Also in the spectator model~\eqref{Eq:define-mean-pT} is not useful,
especially for $e^q(x,p_T)$ where nodes in $p_T$ occur such that
$\la p_T^2\ra$ is negative. In this situation, one gains more 
insight with a different definition of  $\la p_{T,v}^2\ra$ which
is chosen such that one obtains (if it is possible) a useful 
Gaussian approximation of the true $p_T$-dependence at valence-$x$
within a model~\cite{Avakian:2010br}, namely 
\be\label{gauss-model}
     \text{TMD}(x_v,p_T) \approx \text{TMD}(x_v,0) \, 
    e^{-\frac{p_T^2}{\la p_{T,v}^2\ra}}.
\ee
This definition is $x$-dependent, but typically the $x$-dependence 
is weak in the valence-$x$ region~\cite{Avakian:2010br}. For 
definiteness, we choose the value $x_v=0.3$ as reference.
Using this definition, we can directly compare all models, 
see Table~\ref{Table:pT-model}$\,$(b).

\begin{figure}[t!]
\begin{center}
\epsfig{file=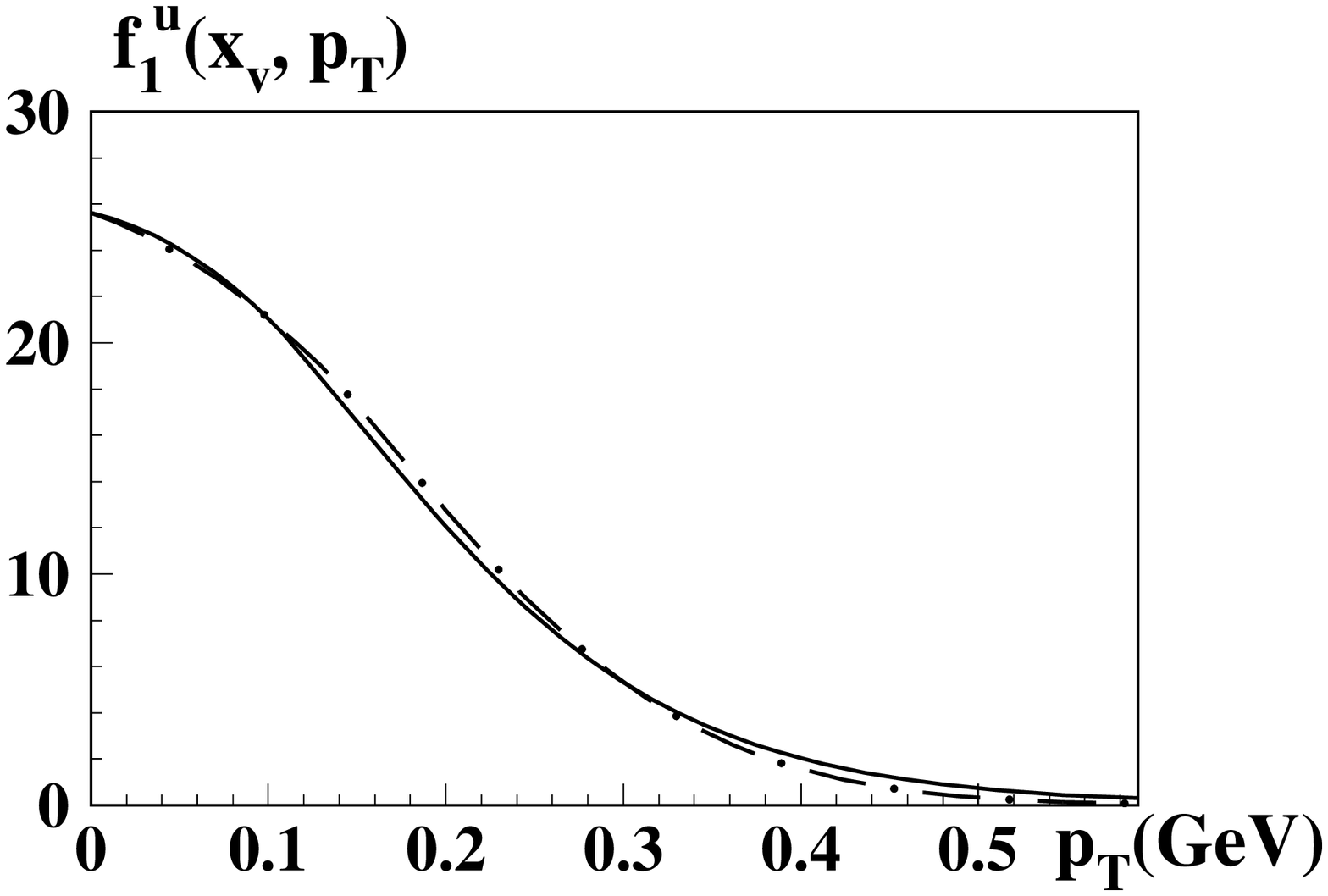,width=0.47\columnwidth}
\epsfig{file=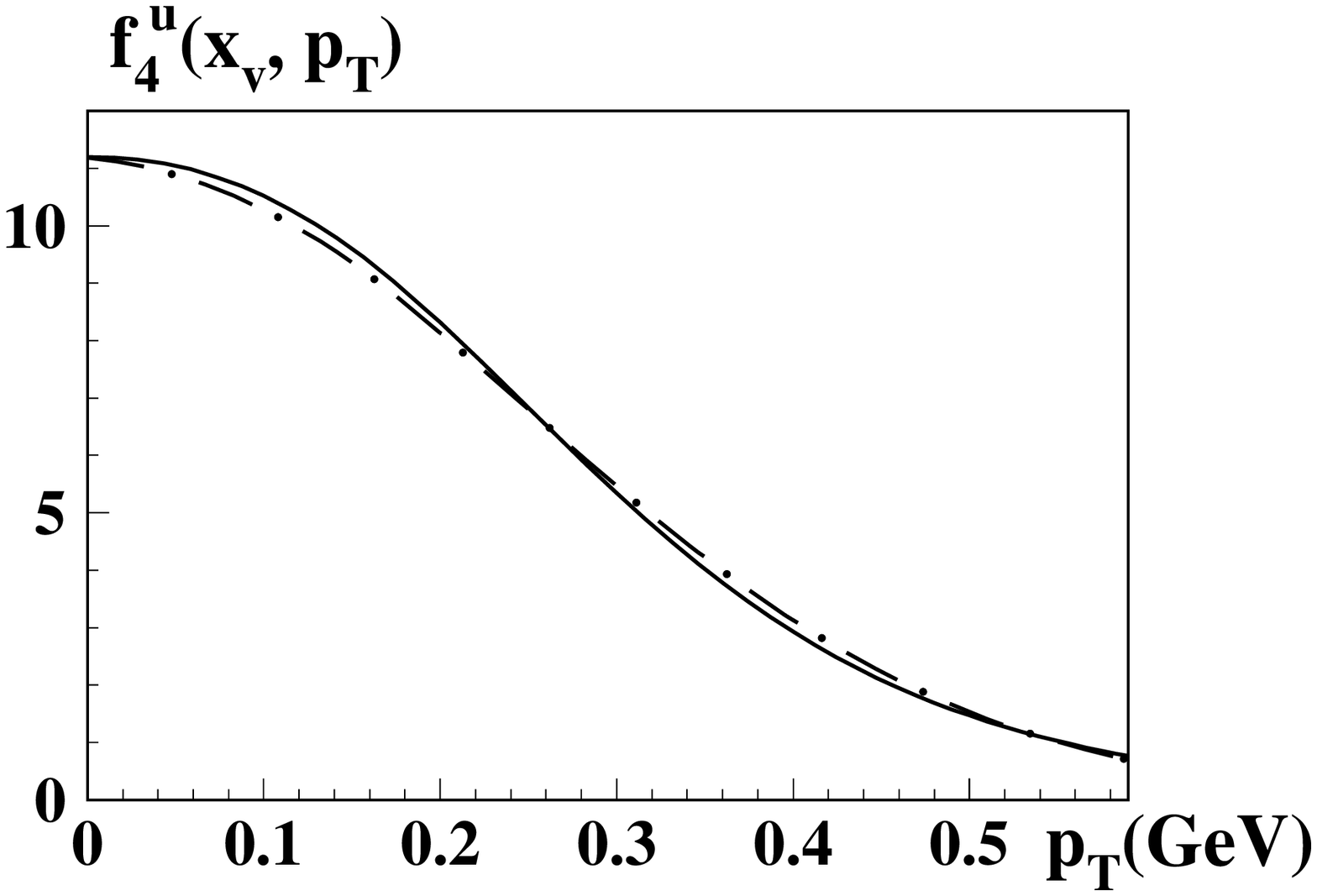,width=0.47\columnwidth}
\end{center}
\vspace{-0.5 truecm}
\caption{\footnotesize{
    $f_1^u(x_v, p_T)$ (left panel) and  $f_4^u(x_v, p_T)$ (right panel) 
    at $x_v=0.3$ as functions of $p_T$. The solid curves show 
    the predictions from the LFCQM, while the dashed-dotted curve are 
    the respective Gaussian approximation from 
    Eq.~\eqref{gauss-model} 
    with the Gauss widths in Table~\ref{Table:pT-model}$\,$(b). }}\label{fig3}
\end{figure}

With the values quoted in Table~\ref{Table:pT-model}$\,$(b) the
true $p_T$-dependence is approximated within 
(5--20)$\%$ depending somewhat on the TMD and model.
As shown in Fig.~\ref{fig3}, in the LFCQM the approximations work 
reasonably well in a large range of $p_T$. However, it is important 
to realize that the TMD picture holds for $p_T^2\ll\mu_0^2$ with the 
initial scale $\mu_0^2\approx 0.176\,$GeV$^2$ in quark models. Thus, 
beyond $p_T^2\gtrsim 2\la p_{T,v}^2\ra$ the non-perturbative
results from quark models for TMDs have no physical meaning. 

The spectator model is the only model with non-trivial flavor 
dependence considered here. Interestingly, the $p_T$-distributions 
of $d$ quarks are systematically broader than those of $u$ quarks. 
The reason behind this is the diquark masses, which set the 
physical scales for the $p_T$-behavior.
The $d$-quark TMDs are given entirely in terms of the heavier 
axial-vector diquark, and are therefore broader. The $u$-quark TMDs 
receive contributions from both scalar and axial-vector diquarks, but 
the lighter scalar diquark dominates which makes the distributions 
narrower.

\section{Comparison to phenomenology}

In order to confront the LFCQM results to phenomenology, it is necessary
to evolve them from the low initial scale to experimentally relevant scales. 
Taking evolution effects into account, the LFCQM as described in the previous 
section, was shown to describe satisfactorily data related to twist-2 TMDs in 
the valence-$x$ region with an accuracy of (10-30)$\%$~\cite{Pasquini:2008ax,
Boffi:2009sh,Pasquini:2010af,Pasquini:2011tk,Pasquini:2014ppa}.
Whether higher-twist TMDs are described with similar success, remains 
to be seen. The recent study~\cite{Courtoy:2014ixa} puts us in the
position to investigate this question for $e^q(x)$. 

For the comparison we will need $e^q(x)$ at a scale of $1.5\,{\rm GeV}^2$.
The pure twist-3 contribution $\tilde{e}^{q}(x)$ follows a complicated
evolution pattern~\cite{Balitsky:1996uh,Belitsky:1997zw,Koike:1997bs} 
typical for subleading-twist distributions, see also the reviews~\cite{Belitsky:1997ay,Kodaira:1998jn}. 
However, in our case $\tilde{e}^{q}(x)=0$ and all we have is 
$x\,e^q(x)=\frac{m_q}{M_N}\,f_1^q(x)$ with the evolution of the latter 
given by the standard evolution of $f_1^q(x)$. To be consistent, we 
also have to make $m_q$ subject to LO evolution of the QCD 
running quark mass (in fact, the quark mass insertion makes the contribution of 
$\frac{m_q}{M_N}\,f_1^q(x)$ ``chiral odd'' and hence a legitimate 
contribution to the chiral odd $e^q(x)$).
It is part of the model, that the value of $m_q$ at the initial 
scale is a sizable constituent quark mass, rather than a small QCD 
current quark mass. But one has to recall that this constituent mass 
has to be understood as an effective parameter describing a quark 
dressed by non-perturbative interactions inside the hadron.

\begin{figure}[t!]
\begin{center}
\epsfig{file=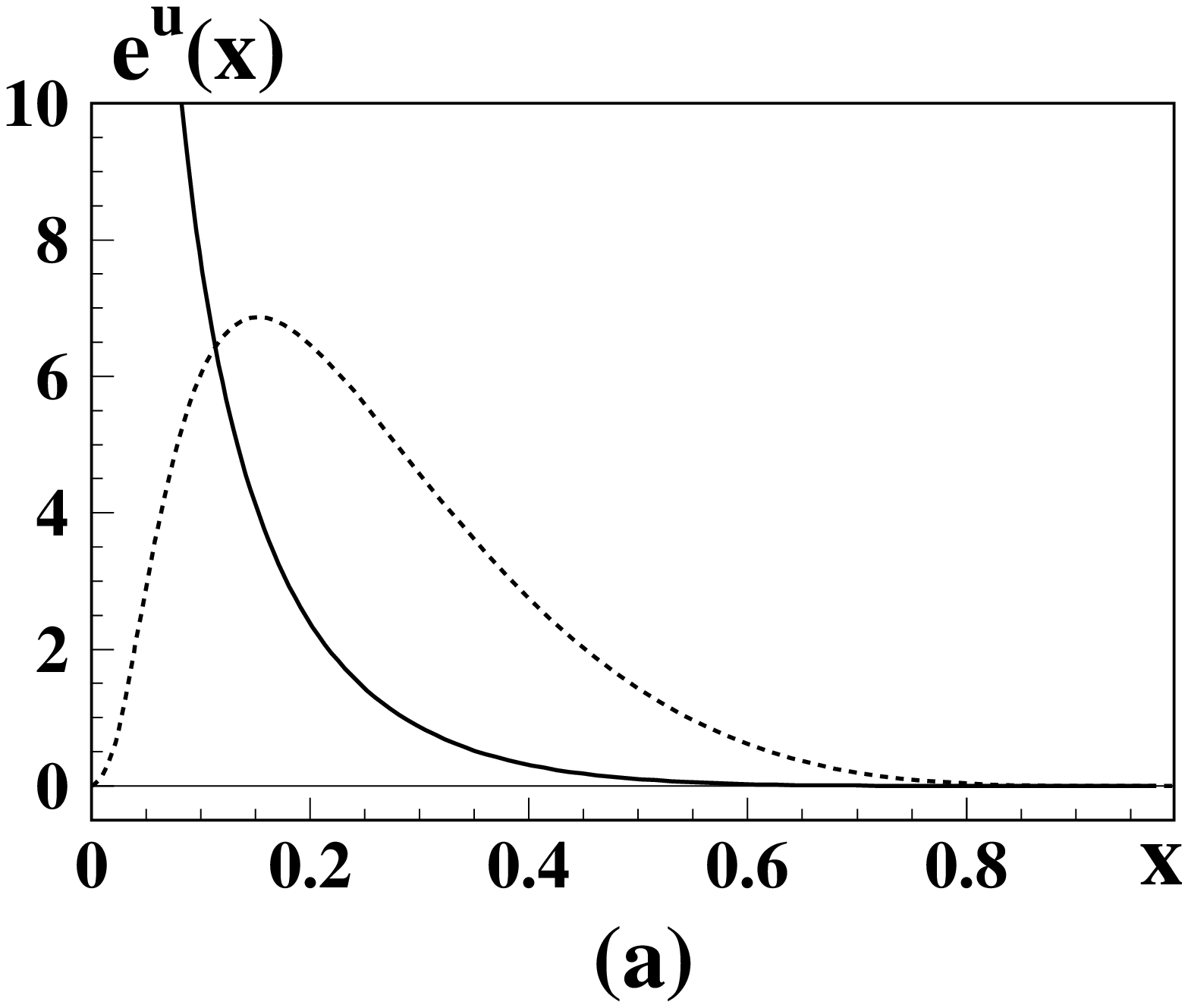,width=0.47\columnwidth}
\epsfig{file=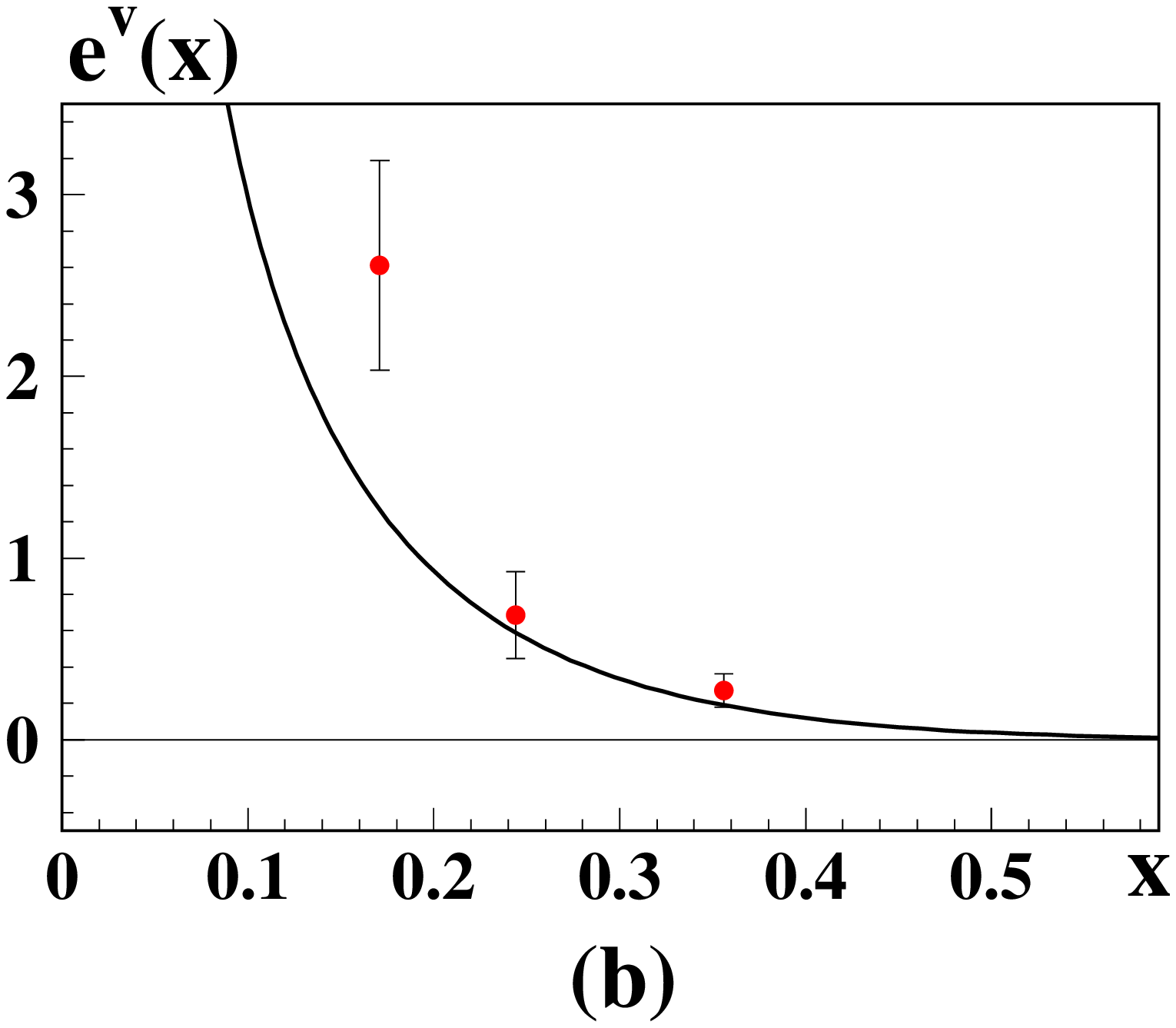,width=0.47\columnwidth}
\end{center}
\vspace{-0.5 truecm}
\caption{\footnotesize{
  (a) The LFCQM predictions for $e^u$ as a function of $x$ at the model scale  $\mu_0^2=0.176$ GeV$^2$ (dashed curve) and LO-evolved to $Q^2=1.5$ GeV$^2$ (solid curve).
 (b) The LFCQM predictions for the combination $e^V=\tfrac{4}{9}\,e^u-\tfrac{1}{9}\,e^d$ as a function of $x$, LO-evolved to $Q^2=1.5$ GeV$^2$ in comparison with the recent extraction 
   of Ref.~\cite{Courtoy:2014ixa}. }}
\label{fig2}
\end{figure}

In Fig.~\ref{fig2} (a) we show $e^u(x)$ at the initial scale of 
$\mu_{0,\rm LO}^2=0.176\,{\rm GeV}^2$, and after LO evolution 
in the above-described way to a final scale of $Q^2=1.5\,{\rm GeV}$
(for technical details of the evolution parameters we refer to~\cite{Pasquini:2011tk}). The results for the $d$-quark 
distribution can be obtained by rescaling by a factor $1/2$  the $u$-quark distribution, according to the $SU(6)$-flavor factors.
Fig.~\ref{fig2} (a) shows that the effects of evolution are sizable, 
and cannot be neglected. The same observations were made also in 
twist-2 case~\cite{Pasquini:2008ax,Boffi:2009sh,
Pasquini:2010af,Pasquini:2011tk,Pasquini:2014ppa}.

Recently the CLAS collaboration has measured azimuthal distributions
of $\pi^+\pi^-$ pairs produced in SIDIS using a longitudinally polarized 6 GeV
electron beam off an unpolarized proton target~\cite{Pisano:2014ila}.
Correlations of final-state 
hadrons~\cite{Efremov:1992pe,Collins:1993kq,Jaffe:1997hf} 
provide a handle to access novel information on the nucleon structure 
in collinear factorization~\cite{Bianconi:1999cd} including $e^q(x)$~\cite{Bacchetta:2003vn}. In this process, one focuses on the 
kinematics where the struck parton fragments into a hadron pair, 
which gives rise to various azimuthal asymmetries.
If we denote by $\sigma^\rightleftarrows$ the cross sections for producing
the hadrons $h_1h_2$ from positive or negative helicity electrons with 
the beam polarization $P_B$ impinging on an unpolarized target, 
$e^\rightleftarrows(l)+N(P)\to e(l^\prime)+h_1(P_{h1})+h_2(P_{h2}) + X$, then 
the observables of interest in our context are~\cite{Bacchetta:2003vn}
\ba\label{Eq:UL-sinphi}
   \frac{1}{2P_B} \;\frac{\di^6\sigma^\rightarrow-\di^6\sigma^\leftarrow}
         {\di^3 u_{hh}\,\di x\,\di y\,\di\phi_R}
   &=& \frac{\alpha^22y\sqrt{1-y}}{2\pi y Q^2}\,{\sin\phi_R}
         \frac{R_T}{Q}\sum\limits_qe_q^2\left(
         \frac{M_N}{m_{hh}} x\,e^q(x)\,H_1^{\sphericalangle q}(u_{hh})
         +\frac1{z_{hh}}\,f_1^q(x)\,\widetilde{G}^{\sphericalangle q}(u_{hh})
         \right), \\
         \label{Eq:UU}
   \frac{1}{2} \;\frac{\di^5\sigma^\rightarrow+\di^5\sigma^\leftarrow}
         {\di^3 u_{hh}\,\di x\,\di y}
   &=&   \frac{\alpha^2(1-y-\frac12y^2)}{2\pi yQ^2}\;
         \sum\limits_qe_q^2\;f_1^q(x)\,D_1^q(u_{hh}),
\ea
where we introduced the abbreviations $u_{hh}\equiv\{z_{hh},\zeta,m_{hh}^2\}$ 
and $\di^3 u_{hh}=\di z_{hh} \di\zeta\di m_{hh}^2$. The DIS variables 
describing lepton scattering are $q=l-l^\prime$, $Q^2=-q^2$, 
$x=Q^2/(2\,P\cdot q)$, and $y=(P\cdot q)/(P\cdot l)$.
The kinematics of the produced hadron pair is described
by the invariant dihadron mass $m_{hh}^2=(P_{h1}+P_{h2})^2$, 
the total longitudinal momentum fraction $z_{hh}=z_1+z_2$ 
transferred from the struck quark to the hadron pair, and 
its relative distribution $\zeta=(z_1-z_2)/z_{hh}$, where
$z_i =(P\cdot P_{hi})/(P\cdot q)$. Finally, $R_T$ is the
component of the relative momentum $\frac12(P_{h1}-P_{h2})$ 
transverse with respect to the total hadron momentum 
$(P_{h1}+P_{h2})$ and given by
$R_T^2 = \frac14(1-\zeta^2)m_{hh}^2 - \frac12(1-\zeta)m_{h1}^2
- \frac12(1+\zeta)m_{h2}^2$.
The angle $\phi_R$ is the inclination of the dihadron plane with 
respect to the lepton scattering plane counted from the direction
of the outgoing lepton~\cite{Bacchetta:2003vn}.

The hadrons $h_1h_2$ can be produced in different relative partial waves, 
and $H_1^{\sphericalangle q}(u_{hh})$ and $\widetilde{G}^{\sphericalangle q}(u_{hh})$
describe the interference of $s$- and $p$-waves~\cite{Bacchetta:2002ux}.
The former is leading twist and arises from the fragmentation
of a transversely polarized quark, the latter is subleading twist
and due to quark-gluon correlations in the fragmentation process. 
In contrast, the leading-twist fragmentation function $D_1^q(u_{hh})$ 
is diagonal in the partial waves.

By deducing information on $D_1^q(u_{hh})$ from the PYTHIA Monte-Carlo 
event generator~\cite{Sjostrand:2003wg} tuned to hadron spectra produced
from $e^+e^-$ collisions in the Belle experiment, and analyzing Belle 
data on azimuthal asymmetries in dihadron production~\cite{Vossen:2011fk}, 
some information on $H_1^{\sphericalangle q}(u_{hh})$ was inferred in~\cite{Courtoy:2012ry}. On the basis of this information, a first 
extraction of $e^q(x)$ from the CLAS data~\cite{Pisano:2014ila} 
was reported in~\cite{Courtoy:2014ixa} (for an earlier attempt to access $e^q(x)$ from SIDIS data
on TMD observables, see~\cite{Efremov:2002ut}).

In Ref.~\cite{Courtoy:2014ixa} it was argued that the CLAS data on 
the ratio of the cross sections~\eqref{Eq:UL-sinphi} and~\eqref{Eq:UU}
cannot be dominated by the second term in~\eqref{Eq:UL-sinphi} 
proportional to $\widetilde{G}^{\sphericalangle q}(u_{hh})$.
Assuming this term to be zero, an approximation referred to as
``WW-scenario'' in~\cite{Courtoy:2014ixa}, yields the extracted
data points for the combination 
$e^V(x)\equiv \frac49(e^u-e^{\bar u})(x)-\frac19(e^d-e^{\bar d})(x)$
shown in Fig.~\ref{fig2} (b) which refers to $Q^2=1.5$ GeV$^2$.

For comparison we show in Fig.~\ref{fig2} (b) the results from the LFCQM 
for the flavor combination
$e^V(x)=\tfrac{4}{9}\,e^u(x)-\tfrac{1}{9}\,e^d(x)$ at the same scale. The agreement with the extraction 
is very satisfactory for the two higher-$x$ bins.
The description of the lowest $x$-bin is less good.
But it is important to recall that the LFCQM is applicable 
in the valence-$x$ region and subject to limitations below 
$x\lesssim 0.2$~\cite{Boffi:2009sh},
\emph{cf.} footnote~\ref{footnote-LFCQM}.

Let us remark that the ``WW scenario'' of Ref.~\cite{Courtoy:2014ixa}
is completely in line with the LFCQM. 
The consistent brute-force neglect of tilde-terms removes not only 
$\widetilde{G}^{\sphericalangle q}(u_{hh})$ but also $\widetilde{e}^{\, q}(x)$.
This is precisely the situation in the LFCQM where
$e^q(x)=\frac{m_q}{M_N}\,f_1^q(x)$ is modeled in terms of 
a sizable constituent quark mass contribution.

It is important to add a cautious remark. The ``WW scenario'' assumed 
in Ref.~\cite{Courtoy:2014ixa} is one possible way of dealing with 
the unknown contribution of $\widetilde{G}^{\sphericalangle q}(u_{hh})$ but 
not the only one. In~\cite{Courtoy:2014ixa} also a ``beyond-WW scenario'' 
was explored where this fragmentation function is allowed to be non-zero, 
with the constraint to reproduce preliminary CLAS data on the double spin 
asymmetry in dihadron production with a longitudinally polarized beam and 
target. This asymmetry is due to $\widetilde{G}^{\sphericalangle q}(u_{hh})$
and compatible with zero within error bars according to the preliminary 
data~\cite{CLAS-double-spin-preliminary}. Although this strongly constrains 
the magnitude of this fragmentation function, a non-zero 
$\widetilde{G}^{\sphericalangle q}(u_{hh})$ compatible with the preliminary 
data~\cite{CLAS-double-spin-preliminary} has a non-negligible impact 
on the extraction of $e^q(x)$. This indicates that the extraction shown 
in Fig.~\ref{fig2} (b) could have sizable unestimated systematic uncertainties. 
The only safe conclusion at the moment is that $e^q(x)$ seems to be non-zero 
in either scenario~\cite{Courtoy:2014ixa}.

Keeping these cautious reservations in mind, we conclude that the
LFCQM prediction for $e^q(x)$ is compatible with the presently
available preliminary CLAS data~\cite{Pisano:2014ila} extracted
in the ``WW-scenario''~\cite{Courtoy:2014ixa} which is 
conceptually in line with the model.

\section{Conclusions}
\label{Sec-8:conclusions}

Sizable azimuthal asymmetries in SIDIS with (un)polarized beams due to 
subleading-twist TMD effects have been observed whose theoretical 
description is not fully clarified. Insights from models can provide 
valuable guidelines.
Quark models in principle offer a tool to evaluate hadronic matrix elements 
of quark-field correlators of any twist~\cite{Jaffe:1991ra}, allowing one 
to model also TMDs, including higher twist. It is therefore of interest 
to explore them as a resource for the interpretation of available data, 
or for predictions for future experiments. For that it is important to 
assess the applicability and limitations of quark models and improve
the understanding of how higher-twist TMDs are modeled. The aim of 
the present work was to contribute to this understanding.

We have shown that exploring the respective equations of motion,
higher-twist TMDs can be decomposed in quark models into contributions 
from leading-twist TMDs, quark-mass terms and pure-interaction dependent 
(``tilde'') terms. This is in some sense analogue to QCD, although the
model interactions are far simpler than the QCD gauge interactions.
Also the meaning of quark mass may differ, as in some models one
may deal with a sizable ``constituent quark mass''.
Nevertheless, the decompositions are fully consistent within the
models, and we have shown that the interaction-dependent tilde-terms 
vanish in formal limits when the model interactions are ``switched off''. 

We have reviewed how this happens in the bag model: the tilde-terms 
vanish when one removes the bag boundary condition~\cite{Jaffe:1991ra}.
Since the latter is designed to ``mimic'' confinement and hence
``gluonic effects'', this demonstrates that the modeling of 
tilde-terms in the bag model is consistent in this sense~\cite{Jaffe:1991ra}.
We also reviewed how tilde-terms arise in the spectator model,
namely due to off-shellness effects~\cite{Jakob:1997wg}.
A new result obtained in this work was the discussion of 
tilde-terms in the chiral quark-soliton model. We have shown that 
these terms vanish if one formally reduces the strength of the 
solitonic field which binds the quarks in that model, and
removes the instanton interactions which ``dress'' the 
light quarks with a dynamically generated mass. These results 
indicate that tilde-terms are ``reasonably'' modeled in these 
approaches and are generated by the respective effective 
interactions.

A remarkable result obtained in this work is the absence of
tilde-terms in the twist-3 TMDs $f^{\perp q}(x,p_T)$ and
$f^{\perp \bar q}(x,p_T)$ in the chiral quark-soliton model. 
Other unpolarized higher-twist TMDs receive significant 
tilde-terms in that model, which arise from the strong chiral 
interactions that bind the quarks in a solitonic field.
For instance, the twist-3 distribution functions $e^q(x)$ 
and $e^{\bar q}(x)$ are, 
in the chiral limit, solely due to a tilde-term which is rather 
sizable in that model~\cite{Schweitzer:2003uy,Wakamatsu:2003uu}.
But in the case of $f^{\perp}$ the chiral interactions
do not induce tilde-terms, and these TMDs are given by
$x\, f^{\perp q}(x,p_T) = f^q_1(x,p_T)$ for 
quarks and   analogous for antiquarks
in the leading order of the large-$N_c$ expansion.
This prediction may have interesting phenomenological
applications.

We also studied models where quarks do not feel explicit 
interactions which, however, not always implies truly 
non-interacting theories. 
In the ensemble of free quarks~\cite{Tangerman:1994eh},
which can be understood as a prototype of more sophisticated
parton model frameworks, the interactions are simply absent 
and the tilde-terms are consequently zero. Parton model approaches 
have important applications, and allow us to separate 
``kinematical'' from ``dynamical'' effects. This leads to 
valuable insights~\cite{Jackson:1989ph,Efremov:2009ze,
D'Alesio:2009kv,Bourrely:2010ng,Anselmino:2011ch},
but does not teach us anything about tilde-terms.

An interesting approach studied in this work in great detail 
is the light-front constituent quark model (LFCQM)
which we extended beyond leading twist.
This approach is based on a light-front Fock-state expansion 
of the nucleon state in terms of on-shell partons ---
each obeying the free EOM. Certain ``unintegrated relations''
among TMDs that are valid in free quark models 
are therefore naturally supported in this model, but not all.
In fact, some free quark model relations among $p_T$-integrated 
TMDs are not supported. One can understand this by recalling 
that the free quark states in the Fock expansion are used
to construct the nucleon light-front wave-function which
encodes non-perturbative information and hence the bound-state nature, through 
certain parameters and the way the 
free quarks states are arranged to form the nucleon state. 
Removing the bound state nature in this case would 
bring us back to the free quark ensemble model.

In order to test the consistency of the different quark model
approaches, we derived a so-called Lorentz-invariance relation
(LIR). Such relations are spoiled in QCD due to gauge interactions,
but they hold in relativistic quark models without gluon degrees
of freedom. We have shown that all quark models satisfy the LIR,
except for the LFCQM. We traced back the reasons
for this to general features of the light-front formalism
which appear at subleading twist~\cite{Burkardt:1991hu}.
The non-compliance of the LFCQM with this specific LIR is
equivalent to the violation of the sum rule for the twist-4
parton distribution function $f^q_4(x)$. In order to satisfy
this sum rule, one has to include light-front zero modes~\cite{Burkardt:1991hu}. An equivalent explanation is
that this sum rule is related to the matrix element of the
minus-component of the electromagnetic current $\la P|J^-|P\ra$.
In a light-front approach, one has to consider overlap contributions 
from higher Fock-state components~\cite{Brodsky:1998hn}. 
Since the modeling of zero-modes or higher Fock-state components
is beyond the scope of the LFCQM, the LIR and the sum rule for 
$f^q_4(x)$ which follows from it, are consequently not supported.

In the LFCQM, where the quarks are non-interacting in the above 
explained sense, tilde-terms are absent and the higher-twist TMDs 
arise from their respective (and in the model consistently described) 
twist-2 contributions and mass terms. Due to the size of the constituent 
quark mass of about 300 MeV in that model, the mass terms are sizable. 
This feature is reasonable and consistent within this model, 
recalling that the results refer to a low renormalization point 
$\mu_0\sim 0.4\,{\rm GeV}$ . 
We presented numerical results from the LFCQM model, and compared 
with other models. 

The LFCQM has been used extensively (more than the other models)
in the past for phenomenological applications in the context of 
leading-twist TMDs, and it was shown that its results are 
compatible with data within a typical model accuracy of about 
(10-30)$\%$. 
A comparison to phenomenology in the twist-3 sector is more difficult, 
as the associated SIDIS observables receive contributions from 4-6  
TMDs and require also a good understanding of presently unknown 
higher-twist fragmentation functions. 

However, recently a phenomenological extraction of the twist-3
parton distribution function $e^q(x)$ was reported~\cite{Courtoy:2014ixa}
based on the collinear interference fragmentation function framework~\cite{Bacchetta:2003vn}. 
Taking into account the evolution from the low initial scale of the LFCQM to the
experimentally relevant scale, we observe a very good agreement
with the extracted result within model accuracy. One should bear
in mind, that the first extraction of information on $e^q(x)$
has unestimated systematic uncertainties~\cite{Courtoy:2014ixa}. 
Nevertheless, the good agreement of the model predictions and the 
phenomenological result is an encouraging indication that the LFCQM 
may be similarly successful in the twist-3 sector as it is in the
twist-2 sector.

Future works will shed more light on the applicability of this
and other quark models to the description of TMDs beyond leading twist, 
and allow us to assess with more confidence to which extent 
quark model approaches are capable to contribute to our
understanding of non-perturbative partonic properties 
at higher twist.

\begin{acknowledgments}
The authors are grateful to A. Bacchetta, A.~Courtoy, S.~Pisano, and C.~Weiss 
for stimulating discussions.
C.L.\ was supported by the Belgian Fund F.R.S.-FNRS \emph{via} 
the contract of Charg\'e de recherches.
P.S.\ was supported by the U.S.~National Science Foundation 
under Contract No. 1406298.
\end{acknowledgments}

\appendix

\section{The new Lorentz-invariance relation}
\label{App:LIR}

Lorentz-invariance relations (LIRs) among TMDs arise when the Lorentz 
decomposition 
of the quark correlator~\eqref{Eq:correlator} contains more TMDs than 
$A_i$-amplitudes~\cite{Tangerman:1994bb}, as is the case in every quark model, 
but not in QCD. We encounter such a situation with the four unpolarized TMDs 
$\{f_1^q,\,e^q,\,f^{\perp q},\,f_4^q\}$ which are given in terms of the three 
amplitudes $\{A_1^q,\,A_2^q,\,A_3^q\}$ in Eqs.~(\ref{Eq:TMD1}--\ref{Eq:TMD4}).
The $A_i$ are Lorentz scalars and can only be functions
of the variables $2\,P\cdot p$ and $p^2$.

Before discussing the derivation of the LIR, we rewrite the expression 
for $f_4^q$ in Eq.~\eqref{Eq:TMD4} as follows
\be
   f^q_4(x,p_T) = \tfrac{1}{2}\, f^q_1(x,p_T) + f^q_{\rm rest}(x,p_T),
   \qquad 
   f^q_{\rm rest}(x,p_T) \equiv 2P^+\int\di p^- \,
                  \frac{(P\cdot p- xM_N^2)}{M_N^2}\,A^q_3.
                  \label{Eq:f4-4e}
\ee
We see that $f^q_4(x,p_T)$ is expressed in terms of $f^q_1(x,p_T)$
and a remaining part $f^q_{\rm rest}(x,p_T)$ related to the amplitude $A_3^q$. The only TMD defined solely in terms
of $A_3^q$ is $f^{\perp q}(x,p_T)$. The goal is therefore to relate
$f^q_{\rm rest}(x,p_T)$ to $f^{\perp q}(x,p_T)$. For that we first
follow Ref.~\cite{Tangerman:1994bb}.

\subsection{Derivation \`a la Tangerman-Mulders}
\label{App:LIR-a-la-T-M}

In this derivation the variables of the amplitude $A_i$ 
are treated as independent quantities. In the next section we will 
see that in quark models the situation can be different.

In order to proceed, we integrate $f^q_{\rm rest}(x,p_T)$ over $p_T$
(in principle, one could formally also take higher transverse moments,
\emph{i.e.} weight by $( p_T^2/2M_N^2)^n$ with $n>1$ before 
$p_T$-integration, though this may raise convergence issues).
Recalling that $p^+=xP^+$, we introduce the convenient variable 
\be\label{Eq:f4-4f}
    \sigma \equiv 2P\cdot p = 2P^+p^-+xM_N^2, \qquad
    \di\sigma = 2P^+\di p^-,
\ee
where the second relation follows for fixed $x$. 
The quark virtuality is then given by
\be\label{Eq:f4-4i}
   p^2 = 2p^+p^--{p}^2_T = x\sigma-x^2M_N^2-{p}_T^2.
\ee
Treating $\sigma$, $x$ and $p^2_T$ as independent variables, we obtain
\begin{align}
 f^q_{\rm rest}(x)
    &=
    \int\di\sigma \int\di^2p_T\, 
    \frac{(\sigma- 2xM_N^2)}{2M_N^2}\,A^q_3(\sigma,x\sigma-x^2M_N^2-{p}_T^2)
    \label{Eq:f4-4j-intermediate}\\
&=
    \int\di\sigma \int\di^2p_T\, 
    \frac{1}{2M_N^2}\,\frac{\di}{\di x}\int^{x\sigma-x^2M_N^2-{p}_T^2}_0\di y\,
    A^q_3(\sigma,y)
    \nonumber\\
&=\frac{\di}{\di x}
    \int\di\sigma \int\di^2p_T\, 
    \frac{1}{2M_N^2}\,\frac{\di p^2_T}{\di p^2_T}\int^{x\sigma-x^2M_N^2-{p}_T^2}_0\di y\,
    A^q_3(\sigma,y)
    \nonumber\\
&=-\frac{\di}{\di x}
    \int\di\sigma \int\di^2p_T\, 
    \frac{p^2_T}{2M_N^2}\,\frac{\di}{\di p^2_T}\int^{x\sigma-x^2M_N^2-{p}_T^2}_0\di y\,
    A^q_3(\sigma,y)
    \label{Eq:f4-4k-intermediate}\\
&=-\frac{\di}{\di x}
    \int\di\sigma \int\di^2p_T\, 
    \frac{p^2_T}{2M_N^2}\,A^q_3(\sigma,x\sigma-x^2M_N^2-{p}_T^2)
    \nonumber\\
  &=
    \frac{\di}{\di x}
    f^{\perp q(1)}(x).
    \label{Eq:f4-4k}
\end{align}
Notice that in the intermediate step~\eqref{Eq:f4-4k-intermediate} 
we integrated by parts with respect to $p_T^2$ which is justified, provided 
$A_3$ falls off at large $p_T$ faster than $1/p_T^4$. This condition also 
ensures that the (1)-moment $f^{\perp q(1)}(x)$ is finite.
Inserting the result~\eqref{Eq:f4-4k} in~\eqref{Eq:f4-4e} yields 
the LIR~\eqref{Eq:LIR-f4}.

\subsection{Derivation for on-shell particles}
\label{App:LIR-on-shell-case}

When the parton (with mass $m_q$) is on-shell as could be the case
in models, then under the $p^-$ integral defining the TMD in terms 
of the amplitude $A_i$, both arguments of $A_i(\sigma,p^2)$ are
fixed in terms of $x$ and $p_T$ 
\be
\begin{aligned}
      p^2 &= m_q^2,\\
      \sigma &= 2\,P\cdot p =xM_N^2+\frac{m_q^2+{p}_T^2}{x}.
\end{aligned}
      \label{Eq:sigma-tau-on-mass-shell}
      \ee
We simulate this situation as follows
\be\label{freeAmpl}
A^q_3(2\,P\cdot p,p^2) = A^{\prime q}_3(2\,P\cdot p)\,\delta(p^2-m_q^2).
\ee
The on-shell condition~\eqref{freeAmpl} allows one to perform the
$p^-$ integration, but it is convenient to refrain from this step. 
Instead, we make use of~\eqref{Eq:sigma-tau-on-mass-shell} 
and \eqref{freeAmpl} under the integral of $f_{\rm rest}^q(x,p_T)$ in 
Eq.~\eqref{Eq:f4-4e} and obtain
\begin{align}
    f^q_{\rm rest}(x,p_T) = 2P^+\int\di p^- \,
    \frac{({p}_T^2+m_q^2-x^2M_N^2)}{2xM_N^2}\,A^q_3
    = \frac{({p}_T^2+m_q^2-x^2M_N^2)}{2xM_N^2}\,f^{\perp q}(x,p_T).
\end{align}
Thus, in contrast to the general case discussed in the previous section, 
here we could complete the task of relating $f^q_{\rm rest}(x,p_T)$ and 
$f^{\perp q}(x,p_T)$ without integrating out transverse momenta. 
Inserting this result in~\eqref{Eq:f4-4e} yields
\be\label{strongLIR}
   f^q_4(x,p_T) = \tfrac{1}{2}\, f^q_1(x,p_T) + 
   \frac{({p}_T^2+m_q^2-x^2M_N^2)}{2xM_N^2}\,f^{\perp q}(x,p_T).
\ee
This relation does not contain new information in ``on-shell'' models,
where it can be derived from the EOM relations. For example, inserting~\eqref{Eq:EOM-f1-fperp} in~\eqref{strongLIR} yields~\eqref{Eq:EOM-f4-f1}.
Nevertheless, we encounter~\eqref{strongLIR} here as an
``unintegrated on-shell version'' of the LIR~\eqref{Eq:LIR-f4}.

Thus, in both on-shell and general cases one finds a relation
expressing $f_4^q$ in terms of $f_1^q$ and $f^{\perp q}$,~\eqref{Eq:LIR-f4} and~\eqref{strongLIR}.
At first glance, these relations seem to be different and this is
puzzling. The essential ingredient of the derivation of LIRs is the 
(unique and complete, in quark models) decomposition~\eqref{Eq:correlator} 
of the correlator in terms of $A_i$ amplitudes, and this is dictated by
Lorentz invariance which all (relativistic) quark models obey. However, 
both versions~\eqref{Eq:LIR-f4} and~\eqref{strongLIR} are {\it formally} 
equivalent (what formally means will become clear shortly).

Starting from the derivative of $f^{\perp(1)q}(x)$ we obtain
\begin{align}
  \frac{\di}{\di x}f^{\perp q(1)}(x)
  &=\frac{\di}{\di x}\left[2P^+\int\di p^-\int\di^2p_T\,\frac{p^2_T}{2M^2_N}
    \,A^{\prime q}_3(2P\cdot p)\,\delta(2p^+p^--p^2_T-m_q^2)\right]\nonumber\\
  &=\frac{\di}{\di x}\int\di^2p_T\,\frac{p^2_T}{2xM^2_N}\,A^{\prime q}_3
    \!\left(xM_N^2+\tfrac{m_q^2+{p}_T^2}{x}\right)\nonumber\\
  &=-\int\di^2p_T\,\frac{p^2_T}{2x^2M^2_N}\left(1-x\,\frac{\di}{\di x}\right)
    A^{\prime q}_3\!\left(xM_N^2+\tfrac{m_q^2+{p}_T^2}{x}\right)\nonumber\\
  &=-\int\di^2p_T\,\frac{p^2_T}{2x^2M^2_N}\,\frac{\di}{\di p^2_T}
    \left[\left(p^2_T+m^2_q-x^2M^2_N\right)A^{\prime q}_3\!
    \left(xM_N^2+\tfrac{m_q^2+{p}_T^2}{x}\right)\right]\nonumber\\
  &=\int\di^2p_T\,\frac{(p^2_T+m^2_q-x^2M^2_N)}{2x^2M^2_N}\,
    A^{\prime q}_3\!\left(xM_N^2+\tfrac{m_q^2+{p}_T^2}{x}\right)
    \label{Eq:surf-1}
    \\
  &=2P^+\int\di p^-\int\di^2p_T\,\frac{(p^2_T+m^2_q-x^2M^2_N)}{2xM^2_N}\,
    A^{\prime q}_3(2P\cdot p)\,\delta(2p^+p^--p^2_T-m_q^2)\nonumber\\
  &=\int\di^2 p_T\, \frac{({p}_T^2+m_q^2-x^2M_N^2)}{2xM_N^2}
    \,f^{\perp q}(x,p_T) \ . 
\end{align}
We again emphasize that in the step leading to~\eqref{Eq:surf-1} 
we integrated by parts, which is legitimate provided $A_3$ falls off at 
large $p_T$ faster than $1/p_T^4$. This condition is anyway required in
order to have a finite results for $f^{\perp(1)q}(x)$.

It is an interesting question to wonder what would happen if $f^{\perp(1)q}(x)$ 
was divergent. In that case, it may (or may not) be possible to introduce 
an appropriate regularization scheme chosen such that $f^{\perp(1)q}(x)$ is 
finite {\it and} the LIR~\eqref{Eq:LIR-f4} is satisfied.
In this context it is interesting to remark, that in the bag model the 
transverse moment $f_1^{(1)q}(x)$ is divergent and needs regularization.
However, $\frac{\di}{\di x}f_1^{(1)q}(x)$ in that model is finite.
In fact, this feature was used in~\cite{Avakian:2010br} to define
the ``regularized''  $f_1^{(1)q}(x)$, namely by integrating 
$\frac{\di}{\di x}f_1^{(1)q}(x)$ (and choosing the integration
constant such that $f_1^{(1)q}(1)=0$).
This in turn implies the interesting possibility that a LIR of
the type~\eqref{Eq:LIR-f4} could hold in a model {\it although}
the associated transverse moment  is undefined.
So far we have not yet encountered such an example.

\subsection{Derivation for spectator model}\label{App-diquark}

In a spectator model, the spectator system is on-shell $m^2_D=(P-p)^2=M^2_N-2P\cdot p+p^2$, and the energy of the struck 
quark is determined by four-momentum conservation. The struck quark 
is off-shell, but both variables of the amplitude $A_i$ are constrained as
\be
\begin{aligned}
      p^2 &= m^2_D - M^2_N + 2P\cdot p,\\
      \sigma &= 2\,P\cdot p = (1+x)M_N^2-\frac{m_D^2+{p}_T^2}{1-x}.
      \label{Eq:sigma-tau-spectator}
\end{aligned}      
\ee 
Like in the free quark model, we can simulate this situation by introducing 
a new amplitude\footnote{
      In Ref.~\cite{Jakob:1997wg} this constraint was formulated as
      $A_i(\sigma,\tau) =A^{\prime\prime }_i(\sigma,\tau) 
      \delta(\tau-\sigma+M_N^2-M_{\rm spectator}^2)$, see Eq.~(68) 
      of Ref.~\cite{Jakob:1997wg}.}
\be\label{freeAmpl-2}
A_3(2P\cdot p,2P\cdot p+m^2_D-M^2_N) = A^{\prime\prime }_3(2P\cdot p)\,
\delta(p^2-2P\cdot p-m^2_D+M^2_N).
\ee
Note that because of the specific form~\eqref{freeAmpl-2}, it is 
actually not necessary to integrate over $p_T$, and we can consider 
directly~\eqref{Eq:f4-4e}. 
If we have four-momentum conservation with the spectator system on-shell, 
we can write
\begin{equation}
    f_{\rm rest}(x,p_T) = 2P^+\int\di p^- \,
    \frac{(1-x)^2M_N^2-({p}_T^2+m_D^2)}{2(1-x)M_N^2}\,A_3=\frac{(1-x)^2M_N^2-({p}_T^2+m_D^2)}{2(1-x)M_N^2}\,f^{\perp }(x,p_T).
\end{equation}
Thus, we obtain the following LIR for spectator models
\be\label{strongLIRSM}
   f_4(x,p_T) = \tfrac{1}{2}\, f_1(x,p_T) + 
   \frac{(1-x)^2M_N^2-({p}_T^2+m_D^2)}{2(1-x)M_N^2}\,f^{\perp }(x,p_T),
\ee
which is satisfied by the separate diquark contributions,
but not by the total result for $f_4^q(x,p_T)$ due to the different
diquark masses. Even if the diquark masses were equal (which is the
case in the limit of a large number of colors), one should notice that 
in contrast to the free quark model, this relation contains the 
model parameter $m_D$. Hence, it is an 
internal model relation, with limited or no validity
beyond the spectator model.

Finally, we are going to check explicitly that the LIR~\eqref{strongLIRSM} 
reduces to~\eqref{Eq:LIR-f4} once integrated over $p_T$. We have
\begin{align}
\tfrac{\di}{\di x}f^{\perp (1)}(x)
  &=\frac{\di}{\di x}\left[2P^+\int\di p^-\int\di^2p_T\,\frac{p^2_T}{2M^2_N}\,
 A^{\prime\prime }_3(2P\cdot p)\,
  \delta(2p^-(p^+-P^+)-2P^-p^+-p^2_T-m^2_D+M^2_N)\right]\nonumber\\
  &=\frac{\di}{\di x}\int\di^2p_T\,\frac{p^2_T}{2(1-x)M^2_N}\,
  A^{\prime\prime }_3\!\left((1+x)M_N^2-\tfrac{m_D^2+{p}_T^2}{1-x}\right)\nonumber\\
  &=\int\di^2p_T\,\frac{p^2_T}{2(1-x)^2M^2_N}\left(1+(1-x)\,\frac{\di}{\di x}\right)
 A^{\prime\prime }_3\!\left((1+x)M_N^2-\tfrac{m_D^2+{p}_T^2}{1-x}\right)\nonumber\\
  &=\int\di^2p_T\,\frac{p^2_T}{2(1-x)^2M^2_N}\,\frac{\di}{\di p^2_T}
  \left[\left(p^2_T+m^2_D-(1-x)^2M^2_N\right)A^{\prime\prime }_3\!
  \left((1+x)M_N^2-\tfrac{m_D^2+{p}_T^2}{1-x}
  \right)\right]\nonumber\\
  &=
 \int\di^2p_T\,\frac{(1-x)^2M_N^2-({p}_T^2+m_D^2)}{2(1-x)^2M^2_N}\,A^{\prime\prime }_3
  \!\left((1+x)M_N^2-\tfrac{m_D^2+{p}_T^2}{1-x}
  \right)\label{Eq:intermediate-step-in-Appendix-on-LIR-in-spectator-model}\\
  &=2P^+\int\di p^-\int\di^2p_T\,\frac{(1-x)^2M_N^2-({p}_T^2+m_D^2)}{2(1-x)M^2_N}\,
  A^{\prime\prime }_3(2P\cdot p)\,\delta( p^2-2P\cdot p-m^2_D+M^2_N)\nonumber\\
  &=\int\di^2 p_T\, \frac{(1-x)^2M_N^2-({p}_T^2+m_D^2)}{2(1-x)M_N^2}\,f^{\perp }(x,p_T).
\end{align}
Also in this model there is a potentially subtle 
step~\eqref{Eq:intermediate-step-in-Appendix-on-LIR-in-spectator-model},
where a total derivative is assumed to vanish. In the model 
of~\cite{Jakob:1997wg} with the choice $\alpha=2$, one has 
$A_3\propto 1/|p_T^2|^{2\alpha-1} \propto 1/p_T^6$.
Consequently, also $f^{\perp q}(x,p_T) \propto 1/p_T^6$. This
ensures the convergence of the transverse moment $f^{\perp(1)q}(x)$,
and the LIR is  satisfied.

\section{\boldmath Sum rule for $f_4(x)$}
\label{App:f4-sum-rule}

The sum rule~\eqref{Eq:f4-sum-rule} for $f^q_4(x)$ can be proven formally in 
several ways, which we shall discuss in App.~\ref{App:sum-rule-proofs}.  Other formal sum rules will be briefly reviewed in 
App.~\ref{App:sum-rule-other-cases}.

\subsection{Proofs of the sum rule for $f^q_4(x)$}
\label{App:sum-rule-proofs}

One way to prove~\eqref{Eq:f4-sum-rule} consists in directly integrating the 
definition~\eqref{Eq:correlator-TMDs4}. It is instructive to do this
in parallel with 
$f_1^q$ defined in~\eqref{Eq:correlator-TMDs1}. Integrating out transverse momenta
in~\eqref{Eq:correlator-TMDs1} and~\eqref{Eq:correlator-TMDs4}, the collinear 
distribution functions are given by
\be
\begin{aligned}
    f_1^q(x)
      &=
   \int \frac{\di z^-}{4\pi} \, 
   e^{i xP^+z^-} \, \la P  | \overline{\psi}(0)\gamma^+
   \psi(z^-) |  P\ra, \\
   \label{Eq:correlator-TMDs4-col}
   \left(\tfrac{M_N}{P^+}\right)^2f_4^q(x)
      &=
   \int \frac{\di z^-}{4\pi} \, 
   e^{i xP^+z^-} \, \la P  |  \overline{\psi}(0)\gamma^-
   \psi(z^-)  |  P\ra,
\end{aligned}
\ee
where we write $\psi^q(z)|_{z^+=z_T^j=0} \equiv \psi^q(z^-)$ for brevity.
Using $2P^+P^-=M^2_N$, the first Mellin moments are 
\begin{align}\label{Eq-app:f4-sum-rule-2}
   2P^+   \int\di x\, f_1^q(x) &= 
   \la P  |  \overline{\psi}(0)\gamma^+\psi(0)  |  P\ra , \nonumber\\
   4P^-\int\di x\, f_4^q(x)& = 
   \la P  |  \overline{\psi}(0)\gamma^-\psi(0)  |  P\ra.
\end{align}
In QCD the electromagnetic current is defined as $J_\mu=\sum_q e_q J_{\mu}^q$ with 
$J_\mu^q= \overline{\psi}(0)\gamma_\mu\psi(0)$. The general decomposition of 
its {\it forward matrix elements} is $\la P|J_\mu^q|P\ra= (2P_\mu) F_1^q(0)$ with 
$F_1^q(0)=N_q$. Thus, from Eq.~\eqref{Eq-app:f4-sum-rule-2} we conclude that
\be\label{Eq-app:f4-sum-rule-3}
  2\int\di x\, f_4^q(x) = \int\di x\, f_1^q(x)= N_q.
\ee
A variant of this proof consists in making use of the fact that Mellin moments 
are Lorentz scalars. Thus, one may go to the nucleon rest frame, where one finds in 
the expressions for the first moments of $f_1^q(x)$ and $f_4^q(x)$ matrix elements 
of the type $\la P|\overline\psi(0)\gamma^\pm\psi(0)|P\ra
=\la P|\psi^\dag(0)(1\pm\gamma^0\gamma^3)\psi(0)|P\ra/\sqrt{2}$.
Now, after the $x$-integration has removed any memory of the light-front direction 
(local matrix element), the contributions 
$\la P|\psi^\dag(0)(\pm\gamma^0\gamma^3)\psi(0)|P\ra$
vanish due to rotational symmetry in the nucleon rest frame implying that
$2\int\di x\, f_4^q(x)$ and $\int\di x\, f_1^q(x)$ are equally normalized.

In quark models, where LIRs are valid, also another formal proof is possible.  
Integrating the LIR~\eqref{Eq:LIR-f4} over $x$, one formally finds
$2\int\di x\, f_4^q(x)=\int\di x\, f_1^q(x)$, since
\be
    \int\limits_{-1}^1\di x \,\frac{\di}{\di x}f^{\perp q(1)}(x) 
    \stackrel{\rm formal}{=} f^{\perp q(1)}(x)\biggl|_{-1}^1 = 
    f^{\perp q(1)}(1)-f^{\perp \bar q(1)}(1) = 0,
\ee
where we used~\eqref{Eq:C-parity} and explored the fact that TMDs vanish 
for $x\to 1$. However, here we tacitly assumed that $f^{\perp q(1)}(x)$ is a 
continuous function of $x$ including the point $x=0$. This can, but does 
not need, to be the case in models. Thus, in the general case one could 
find that the small $x$-behavior invalidates this proof, due to
\be
    \int\limits_{-1}^1\di x \,\frac{\di}{\di x}
    f^{\perp q(1)}(x) = \lim\limits_{\epsilon\to 0}
    \left(-f^{\perp q(1)}(\epsilon)+f^{\perp \bar q(1)}(\epsilon)\right) \neq 0.
\ee
A gaze at models provides intuition. In both the bag model and $\chi$QSM~\eqref{Eq:f4-sum-rule} is satisfied, which is straightforward to check by
directly integrating model expressions and exploring rotational (in bag model)
or hedgehog (in $\chi$QSM) symmetries. 
In the bag model (where one has to keep in mind the reservations due to the 
unphysical negative-$x$ region, see footnote~\ref{footnote-bag-model}),
$f^{\perp q(1)}(x)$ is a continuous function at $x=0$, so one can also integrate 
the LIR to prove~\eqref{Eq:f4-sum-rule}.
But in the $\chi$QSM, which describes at $x<0$ physical TMDs according to~\eqref{Eq:C-parity}, one has $x\,f^{\perp q(1)}(x)=f_1^q(x)$ and the latter exhibits 
a discontinuity at $x=0$ that ensures positivity~\cite{Diakonov:1996sr}. 
Thus, in the $\chi$QSM the sum rule~\eqref{Eq:f4-sum-rule} is valid, but 
cannot be proven by integrating the LIR. 

As the last proof in quark models, we notice that $f^q_{\rm rest}(x)$ in the intermediate step 
\eqref{Eq:f4-4j-intermediate} can be rewritten as 
\cite{Tangerman:1994bb}
\be\label{Eq:XXX-6}
    f^q_{\rm rest}(x) =
    \int\di\sigma\int\di\tau\int\di^2p_T\, 
    \frac{(\sigma- 2xM_N^2)}{2M_N^2}\;
        \delta(x\sigma-x^2M_N^2-{p}_T^2-\tau)
    \;A^q_3(\sigma,\tau).
\ee
Integrating this expression over $x$ we obtain
\be\label{Eq:XXX-7}
    \int\limits_{-1}^1\di x\,
    f^q_{\rm rest}(x) 
    \stackrel{\rm formal}{=}
    \int\di\sigma\int\di\tau\int\di^2p_T\, 
    \frac{A^q_3(\sigma,\tau)}{2M_N^2}\;
    \int\limits_{-1}^1\di x\,(\sigma- 2xM_N^2)\,
    \delta(x\sigma-x^2M_N^2-{p}_T^2-\tau) = 0,
\ee
which vanishes because we deal with an integral of the type 
\begin{align}\label{Eq:XXX-8}
   \int\di x \,v^\prime(x)\,\delta(v(x)) =
   \int\di x \sum_i v^\prime(x)\;\frac{\delta(x-x_i)}{|v^\prime(x)|}
   = \sum_i {\rm sign}\,v^\prime(x_i),
\end{align}
where the $x_i$ are simple zeros of the argument $v(x)$, and
our function $v(x)=x\sigma-x^2M_N^2-{p}_T^2-\tau$ is such that
$v^\prime(x_{1,2})=\mp\sqrt{\sigma^2-4M_N^2(\tau+{p}_T^2)}$.
Using this expression for $v'(x)$ in~\eqref{Eq:XXX-8}, one 
formally finds that $\int\di x\,f^q_{\rm rest}(x) = 0$, 
confirming~\eqref{Eq:XXX-7} and proving the sum rule~\eqref{Eq:f4-sum-rule}. 
However, in a specific model one has to investigate carefully whether 
$x_i\in [-1,1]$ such that the integrated $\delta(v(x))$ 
has indeed support in the integration region.

\subsection{Other potentially violated sum rules} 
\label{App:sum-rule-other-cases}

\

Sum rules like~\eqref{Eq:f4-sum-rule} are referred to as formal. They are 
mathematically correct. But in the formal theoretical evaluation of
such a sum rule, a $\delta(x)$-singularity (if present) is integrated over, 
and contributes to the result. 
However, the experimental test of such a sum rule will only include results 
inferred (and extrapolated) from data taken at finite $x>0$. Hereby of course 
the contribution of the $\delta(x)$-singularity will be missed, 
and sum rule perceived as violated.

We are not aware of how (even in principle) the twist-4 sum 
rule~\eqref{Eq:f4-sum-rule} could be tested, but there are other 
sum rules which can be tested experimentally.
The most famous example is the long-discussed and still unsettled possible 
violation of the Burkhardt-Cottingham sum rule which features the twist-3
parton distribution function $g_T^q(x)$~\cite{Burkhardt:1970ti}.
Also the sum rule of the twist-3 parton distribution function $h_L^q(x)$ 
was debated~\cite{Burkardt:1995ts}. 

But the most interesting case in the context of this work is the Jaffe-Ji 
sum rule~\cite{Jaffe:1991ra} connecting $e^q(x)$ to the pion-nucleon 
sigma-term $\sigma_{\pi N}$. By exploring QCD equations of motion, $e^q(x)$ 
can be decomposed as follows~\cite{Efremov:2002qh} 
\be\label{Eq:e-decomposition}
   e^q(x) = \frac{\delta(x)}{2M_N}\, \la P | \overline\psi(0)\psi(0)  |  P\ra 
          + \tilde{e}^{q}(x) + e_{\rm mass}^q(x).
\ee
Here $\tilde{e}^q(x)$ and $e_{\rm mass}^q(x)$ denote, respectively,
the pure twist-3 and mass term, which in QCD have the properties 
\be\label{Eq:e-mas-tilde-sum-rules}
 \int\di x\,\tilde{e}^q(x)=\int\di x\,x\,\tilde{e}^q(x)=
    \int\di x\,e_{\rm mass}^q(x)=0.
\ee 
For $x\neq 0$ the mass term is expressed in QCD as well as in quark models by 
$xe_{\rm mass}^q(x)=\frac{m_q}{M_N}\,f_1^q(x)$. 
Thus, in QCD the sum rule (disregarding a small doubly isospin violating term)
for $e^q(x)$ is given by
\be\label{Eq:e-sigma-piN-sum-rule}
   \sum\limits_{q=u,d} \int\di x\,e^q(x) = \frac{\sigma_{\pi N}}{m}, \qquad
   m=\tfrac{1}{2}(m_u+m_d).
\ee
A $\delta(x)$-contribution in $e^q(x)$ was found in $(1+1)$-dimensional models~\cite{Burkardt:1995ts}, perturbative one-loop light-front calculations~\cite{Burkardt:2001iy}, and non-perturbative calculations in the $\chi$QSM~\cite{Schweitzer:2003uy,Wakamatsu:2003uu}. 
In the one-loop dressed-quark model of~\cite{Burkardt:2001iy}, $\delta(x)$ 
emerged as a $p^+$ zero mode in light-front time-ordered perturbation theory. 
In the $\chi$QSM, the coefficient of the $\delta(x)$-function (and hence 
$\sigma_{\pi N}$, see~\cite{Schweitzer:2003sb}) is related to the quark vacuum 
condensate~\cite{Schweitzer:2003uy,Wakamatsu:2003uu}, a quantity with central 
importance as order parameter of spontaneous chiral breaking.

No $\delta(x)$ singularity appears in the bag~\cite{Jaffe:1991ra,Signal:1996ct}
or spectator~\cite{Jakob:1997wg} models. Particularly interesting in our
context is the model with massive quarks in light-front one-loop Hamiltonian 
perturbation theory with light-front gauge~\cite{Mukherjee:2009uy} where also 
no $\delta(x)$
contribution was found (this was in fact  impossible, because in contrast 
to~\cite{Burkardt:2001iy}, in the calculation of~\cite{Mukherjee:2009uy} a 
prescription for the operator $\frac{1}{\partial^+}$ was chosen, which discards
$p^+$ zero modes). The $\tilde{e}^{q}(x)$ and $e_{\rm mass}^q(x)$ from~\cite{Mukherjee:2009uy} do not satisfy~\eqref{Eq:e-mas-tilde-sum-rules}. 
However, remarkably $\int\di x\,(\tilde{e}^{q}+e_{\rm mass}^q)(x)$ 
nevertheless satisfies the sum rule~\eqref{Eq:e-sigma-piN-sum-rule}.
Thus, in this calculation the information on $\sigma_{\pi N}$ is, instead
of being concentrated in the point $x=0$, redistributed over the whole
interval $0 < x < 1$. The same kind of ``holographic principle'' 
is observed in the LFCQM, see Sec.~\ref{Sec-X:numerical-results}.

\section{\boldmath $f^\perp(x,p_T)$ in the chiral quark soliton model}

The derivation of the $\chi$QSM expression for $f^\perp(x,p_T)$ proceeds 
analogously to the calculation of the unpolarized and helicity 
TMDs~\cite{Schweitzer:2012hh}.
The result for the flavor combination $u+d$, which is leading in 
the large-$N_c$ limit, is 
(the result for the flavor combination $u+d$ of $f_1$ 
from \cite{Schweitzer:2012hh} is included for reference)
\begin{align}
    p_T^i \, f^\perp(x,p_T) &= N_c M_N^2 \sum\limits_{n,\,\rm occ}
    \phi^\ast_n(\vec{p})\gamma^0\gamma_T^i\phi_n(\vec{p})\bigl|_{p^3=xM_N-E_n} ,\\
                f_1(x,p_T) &= N_c M_N \sum\limits_{n,\,\rm occ}
    \phi^\ast_n(\vec{p})(1+\gamma^0\gamma^3)\phi_n(\vec{p})\bigl|_{p^3=xM_N-E_n} .
\end{align}
Notice that for $x<0$, these formulae describe $(-f_1)$ and $(+f^\perp)$ distributions for
antiquarks. 

In the main text we have proven the remarkable result 
that $x\,f^{\perp q}(x,p_T) = f_1^q(x,p_T)$, see 
Sec.~\ref{Sec-7:eom-in-interacting-models}. 
The proof given in Sec.~\ref{Sec-7:eom-in-interacting-models}
was formal: it explored the model EOM, but was formulated in terms 
of the general correlator expression. Here as a double-check, we present 
a proof formulated in terms of the single-quark wave-functions, i.e.\
with the  general correlator evaluated using the techniques of 
Ref.~\cite{Schweitzer:2012hh}.

Working in the chiral limit $m_q\to 0$ and in momentum 
space, the Hamiltonian is given by 
$H=\gamma^0\vec\gamma\cdot\vec{p}+\gamma^0M\,U^{\gamma_5}$.
The single-quark wave-functions satisfy
$H\phi_n(\vec{p})=E_n\phi_n(\vec{p})$, and we have
the obvious identity 
\begin{align}
 0 &= \phi^\ast_n(\vec{p})\left[
   (E_n-H)(\gamma^0\gamma_T^j)(1+\gamma^0\gamma^3)+
   (1+\gamma^0\gamma^3)(\gamma^0\gamma_T^j)
   (E_n-H)\right]\phi_n(\vec{p})\nonumber\\
   &= 2\,\phi^\ast_n(\vec{p})\left[
   (E_n+p^3)\gamma^0\gamma_T^j-p_T^j( 1+\gamma^0\gamma^3)\right]\phi_n(\vec{p}), 
\end{align}
where the second step follows after a little Dirac algebra.
In order to apply this result to TMDs, we include the prefactor $N_cM_N$, 
sum over occupied quark levels, and introduce the constraint $p^3=xM_N-E_n$ 
which allows us to replace $(E_n+p^3)$ by $xM_N$. As a result, we obtain
\be
   0 = N_cM_N\sum\limits_{n,\,\rm occ}\phi^\ast_n(\vec{p})\left[
       xM_N\gamma^0\gamma_T^j-p_T^j(1+\gamma^0\gamma^3)\right]\phi_n(\vec{p}) 
     = p_T^j\left[x\, f^\perp(x,p_T) - f_1(x,p_T)\right],
\ee
which proves that $x\, f^\perp(x,p_T) = f_1(x,p_T)$. 
Two remarks are in order. First, we see that the relation is satisfied 
for each quark level separately. This is so, because we used the 
EOM for the single quark states. Second, we see explicitly the off-shellness 
of the quark in the $n^{\rm th}$ level $p_n=(E_n,\bm{p}_T,xM_N-E_n)$, namely
$p^2_n=E_n^2-(xM_N-E_n)^2- p_T^2\neq0$, which would have been expected for 
massless on-shell quarks.

\end{document}